\documentclass{iopart}

\usepackage{epsfig}
\usepackage{amscd,amssymb}
\usepackage{graphicx,xspace}
\usepackage{pstricks}
\usepackage[left=3cm,top=3cm,right=3cm,bottom=3cm]{geometry}
\usepackage{color}

\def\be{\begin{equation}}
\def\ee{\end{equation}}

\def\bea{\begin{eqnarray}}
\def\eea{\end{eqnarray}}

\def\e{\epsilon}

\def\t{\tau}
\def\tz{{\tau_0}}

\newcommand{\hf}[4]{\mathop{{_2}F_1}\left(#1,#2,#3;#4\right)}

\begin{document}

\title[Time evolution of 1D gapless models]{Time evolution of 1D gapless models
  from a domain-wall initial state: SLE continued? }

\author{Pasquale Calabrese${}^{1}$, Christian Hagendorf${}^{2}$,
and Pierre Le Doussal${}^{2}$}

\address{$^{1}$Dipartimento di Fisica dell'Universit\`a di Pisa and INFN,
Pisa, Italy}

\address{$^{2}$ CNRS-Laboratoire de Physique Theorique de l'Ecole Normale
Superieure, 24 rue Lhomond, 75231 Paris, France}

\begin{abstract}
We study the time evolution of quantum one-dimensional gapless systems
evolving from initial states with a domain-wall. 
We generalize the path-integral imaginary time approach that together with
boundary conformal field theory allows to derive the time and space dependence 
of general correlation functions. 
The latter are explicitly obtained for the Ising universality class, and the 
typical behavior of one- and two-point functions is derived for the general 
case. 
Possible connections with the stochastic Loewner evolution are discussed and 
explicit results for one-point time dependent averages are obtained for
generic $\kappa$ for boundary conditions corresponding to SLE.
We use this set of results to predict the time evolution of the entanglement
entropy and obtain the universal constant shift due to the presence of a 
domain wall in the initial state.
\end{abstract}

\maketitle

\section{Introduction}

The experimental realization of trapped quasi-one-dimensional atomic gases
during the last few years
\cite{gv-01,gb-01,mske-03,sm-04,pw-04,kwt-04,kwt-05,aewkd-08,bdw-07}
has provided a new impetus for the study of
the effects of strong correlations on the physical properties of fundamental
quantum mechanical systems of interacting particles.
One very interesting aspect of this still emerging field, in contrast to
traditional condensed matter physics, is that
it is possible to follow the unitary time evolution
in the absence of any dissipation \cite{uc,uc2,uc3,spinor,kww-06,wt-07,tc-07}.

The absence of a general theoretical framework to investigate the
non-equilibrium dynamics of extended quantum systems led to the study of many
one-dimensional models whose dynamics can be solved exactly (see e.g. Refs.
\cite{ao-03,aop-03,sps-04,cc-05,dmcf-06,cl-05,mg-05,cc-06,gg,c-06,rmo-06,cdeo-07,gdlp-07,cc-07,ep-07,klr-07,cc-07b,rdo-07,bpz-07,gp-07,ekpp-07,bs-07,brd-08,qzm,ep-08},
but the list is far from being exhaustive) and inspired the
developing of new numerical methods among which time-dependent
density-matrix renormalization-group (DMRG) is the most successful
\cite{tdmrg}, giving the first opportunity to investigate the out of
equilibrium behavior also of non-integrable systems as e.g. in Refs.
\cite{dmcf-06,ksdz-05,ksz-05,kla-06,rmrnm-06,mwnm-06,kkmgs-07,hrmfd-08,lk-08,arf-08}.

>From the study of specific models it is difficult to draw general conclusions
on the non-equilibrium physics of quantum systems. Hence it is desirable to
have frameworks that, at least in some relevant time windows, can give results
valid for a large class of systems. In Refs. \cite{cc-06,cc-07} a method to
tackle the time evolution after a quantum quench (i.e. a sudden change of a
Hamiltonian parameter) has been proposed.
Within this method the real-time path integral representation of a
$d$-dimensional quantum system is reduced to the thermodynamics of a $d+1$
dimensional classical system in a slab geometry.
The time translation invariance is explicitely broken by the initial state that
plays the role of a boundary state in the slab. When the Hamiltonian $H$
governing the time evolution is at or close to a quantum critical point, one
can use the renormalization group (RG) theory of boundary critical phenomena
(see e.g. \cite{dd}) to study the regime of time and length scales much larger
than microscopic ones.
In Refs. \cite{cc-06,cc-07} this approach has been mainly applied to the time
evolution of one-dimensional (1D) systems since the 1+1 dimensional strip is
described asymptotically by a boundary conformal field theory (CFT).

Two very intriguing properties were derived by using this approach. First, connected correlations start forming
only after two points are causally connected (horizon effect). Second, the large time asymptotic of correlation
functions are the same as in a thermal state with effective temperature $\beta_{\rm eff}=4\tz$ ($\tz$ being
related to the inverse of the mass gap in the initial state, see below), despite the fact that the system is
always in a pure state. These predictions have been confirmed by a number of exact calculations in specific
models. This path-integral approach is generalizable to several different situations and it is very surprising,
despite the success in the simplest case, that this has been done only in a few cases (in Ref. \cite{cc-07b} for
the so called local quench and in Ref. \cite{gc-08} for the evolution of the
order-parameter in higher dimensional systems).

In this paper we apply this imaginary time path-integral to the case in which the initial state is not
translationally invariant but contains a domain-wall. We study the resulting time evolution governed by a
gapless Hamiltonian using boundary CFT. A part from the per se interest, there are at least two additional
reasons to study such an initial state. On one hand, the presence of a domain-wall generates a non-trivial
transport, as shown by the exact solution in XX and XY chains \cite{kink}. On the other hand, in 2D classical
critical systems, these are the typical boundary conditions used to set up the stochastic Loewner evolution
(SLE), a mathematical rigorous approach to describe stochastic and geometric properties of 2D critical systems.
Hence one must ask the question of whether SLE could be used to predict the time evolution of these quantum
models.

The paper is organized as follows. In Sec. \ref{sec2} we review the imaginary-time path-integral approach to
quantum quenches. We specifically show how to tackle the case of a domain-wall in the initial state by means of
CFT. In Sec. \ref{sec3} we apply the method to the Ising model for which all the relevant correlation functions
in the upper half-plane have been calculated in Ref. \cite{bx-90}. In Sec. \ref{sec4} we describe which features
of this quenches can be extracted by general CFT scaling without the knowledge of the full correlators in the
upper half-plane and then in the following section \ref{sec5} these results are used to give prediction for the
time-evolution of the entanglement entropy. In Sec. \ref{sec6} we argue about
possible connections with SLE and we discuss the case of two domain-walls in
the initial state. 
Finally in Sec. \ref{sec7} we critically discuss our findings and open problems. In \ref{appA} we report some
useful formulas to perform the analytic continuations and in \ref{appB} we derive the time evolution for an
homogeneous quench of the two-point functions of different operators.

\section{The CFT approach to quantum quenches}
\label{sec2}

In Refs. \cite{cc-06,cc-07} it has been shown how to extract from path-integral the unitary time-evolution of a
$d$-dimensional system prepared in a state $|\psi_0 \rangle$ that is not an eigenstate of $H$. The expectation
value of a local operator ${\cal O}({\bf r})$ at time $t$ is \be \langle {\cal O}(t,{\bf r})\rangle= Z^{-1}
\langle \psi_0 | e^{i H t-\e H} {\cal O}({\bf r}) e^{-i H t-\e H}| \psi_0 \rangle\,, \label{Oexp} \ee
where the damping factors $e^{-\e H}$ have been introduced in order to
make the path-integral representation of the expectation value absolutely convergent. $Z=\langle\psi_0 |e^{-2\e
H}|\psi_0 \rangle$ ensures that the expectation value of the identity is one. Eq. (\ref{Oexp}) may be
represented by an analytically continued path integral in imaginary time, described by the evolution operator
$e^{- (\tau_2-\tau_1) H}$, which takes the boundary values $\psi_0$ on $\tau=\tau_1=-\e-it$ and on
$\tau=\tau_2=\e-it$. The operator ${\cal O}$ is inserted at $\tau=0$. The width of the slab is $2\e$. $\tau_1$
and $\tau_2$ must be considered as real numbers during all the calculation. Only at the end they are continued
to their effective values $\pm\e-i t$. The real-time non-equilibrium evolution of a $d$ dimensional system is
then reduced to the thermodynamics of a $d+1$ field theory in a slab geometry with the initial state $|\psi_0
\rangle$ playing the role of boundary condition at both the borders of the slab. More generally the operator
${\cal O}$ in Eq. (\ref{Oexp}) can depend on several points, and in the $d=1$ case on which we focus below, we
denote $r_i$ the set of these points, together with possible domain-wall positions in the initial state.

We wish to study this setting in the limit where $t$ and all separations $|r_i-r_j|$ are much larger than any
microscopic length and time scales, so that renormalization group (RG) theory can be applied. If $H$ is at or
close to a quantum critical point, the bulk properties of the critical theory are described by a bulk RG fixed
point (or some relevant perturbation thereof). Any {\it translationally invariant} boundary condition flows to
one of a number of possible boundary fixed points \cite{c-84,c-05}, and we may replace $|\psi_0\rangle$ by the
appropriate RG-invariant boundary state $|\psi_0^*\rangle$ to which it flows. The difference may be taken into
account by assuming that the RG-invariant boundary conditions are not imposed at $\tau=\tau_1$ and $\tau_2$ but
at $\tau=\tau_1-\tau_0$ and $\tau=\tau_2+\tau_0$. $\tau_0$ is given by the absolute value of the extrapolation
length and it characterizes the RG distance of the actual boundary state from the RG-invariant one \cite{dd}. In
the quantum non-equilibrium problem, $\tau_0$ is expected to be of the order of the correlation length in the
initial state, that is the inverse energy gap \cite{cc-06,cc-07}. The effect of introducing $\tau_0$ is simply
to replace $\e$ by $\e+\tau_0$. The limit $\e\to0^+$ can now safely be taken, so the effective width of the slab
results to be $2\tau_0$. For simplicity in the calculations, in the following we will consider the equivalent
slab geometry between $\tau=0$ and $\tau =2\tau_0$ with the operator ${\cal O}$ inserted at $\tau=\t_0+it$.

If the initial condition (and so the boundary one in the imaginary time representation) is not translational
invariant, the approach must be changed accordingly. For example, in Ref. \cite{cc-07b} it has been shown how to
deal with initial conditions corresponding to the joining of two semi-infinite ground-states of a critical
Hamiltonian. The case of domain-walls in the initial states, in which we are interested in this paper, is also
easily tackled with boundary CFT. In fact, as shown by Cardy \cite{c-84}, any domain-wall at some position $r_i$
can be viewed as the insertion of an appropriate operator $\Phi_{ab}$ at position $r_i$. The label $a$
stands for the boundary condition to the right, and $b$ to the left of the domain-wall respectively. $\Phi_{ab}$ is
called {\it boundary condition
  changing} operator for obvious reasons.
For any bulk CFT, there are a number of possible boundary states and
consequently a number of boundary condition changing operators between them
that are classified and their scaling dimensions are known (see e.g.
\cite{c-84}) in the most relevant cases.

In the upper half-plane with a boundary condition having $m$ domain-walls (including the
point at $\infty$) on the real axis, any $n$-point function can be obtained in terms of a $(2n+m)$-point
correlation of the $n$ operators plus $m$ boundary condition changing ones. The $2n$ comes from the bulk points
and their images. Increasing $n$ and $m$ technical difficulties to get these correlations increase considerably.
For this reason, we focus on one- and two-point functions with two domain-walls (one at $\infty$). The
difficulty of this calculation corresponds to that of four- and six-point functions in the complex plane.

\subsection{The general strategy by conformal mapping}

\begin{figure}[htpb]
  \centering
  \begin{pspicture}(0,0.5)(7,5)
    \psframe[fillstyle=solid,fillcolor=lightgray,linewidth=0](1,1.3)(7,1.5)
    \psframe[fillstyle=hlines,fillcolor=lightgray,linewidth=0](1,1.3)(4,1.5)
    \psframe[fillstyle=vlines,fillcolor=lightgray,linewidth=0](4,1.3)(7,1.5)
    \psframe[fillstyle=solid,fillcolor=lightgray,linewidth=0](1,4)(7,4.2)
    \psframe[fillstyle=hlines,fillcolor=lightgray,linewidth=0](1,4)(4,4.2)
    \psframe[fillstyle=vlines,fillcolor=lightgray,linewidth=0](4,4)(7,4.2)
    \psline[linewidth=1.5pt](1,1.5)(7,1.5)
    \psline[linewidth=1.5pt](1,4)(7,4)
    \psline[linestyle=dotted](0.8,2.75)(7,2.75)
    \psline{->}(0.8,1.3)(0.8,4.5)
    \psline{->}(1,0.8)(7,0.8)
    \psline(4,0.7)(4,0.9)
    \rput[c](0.5,1.5){$0$}\rput[c](0.5,2.75){$i\tau_0$}\rput[c](0.4,4){$2i\tau_0$}
    \rput[c](0.5,4.75){$i\tau$}
    \psline(0.75,1.5)(0.85,1.5)\psline(0.75,2.75)(0.85,2.75)\psline(0.75,4)(0.85,4)
    \rput(4,1.1){$0$}\rput(6.9,1.1){$r$}
    \rput(2.5,1.1){$-$} \rput(5.5,1.1){$+$}
    \rput(2.5,4.4){$-$} \rput(5.5,4.4){$+$}
    \psdot(2,2.75)\rput[c](2.3,3.2){$w_1=r_1+i\tau_1$}
    \psdot(5,2.75)\rput[c](5.7,3.2){$w_2=r_2+i\tau_2$}
    \rput(1.5,1.8){\Large ${\mathtt S}$}
  \end{pspicture}\hspace{1.cm}
  \begin{pspicture}(7,5)
    \psframe[fillstyle=solid,fillcolor=lightgray,linewidth=0](0,0.8)(7,1)
    \psframe[fillstyle=hlines,fillcolor=lightgray,linewidth=0](0,0.8)(3.5,1)
    \psframe[fillstyle=vlines,fillcolor=lightgray,linewidth=0](3.5,0.8)(7,1)
    \psline[linewidth=1.5pt](0,1)(7,1)
    \psline[linestyle=dashed,linecolor=gray](3.5,1)(1,2.64)
    \psline[linestyle=dashed,linecolor=gray](3.5,1)(4,3.95)

    \rput(1.75,0.5){$\mathbf{-}$}\rput(5.25,0.5){$\mathbf{+}$}
    \rput(2.4,3.45){$\rho$}
    \psdot(1,2.64)\psdot(4,3.95)\psline{<->}(1,2.64)(4,3.95)
    \psarc[linestyle=dotted](3.5,1){3}{0}{180}

 \rput(4,4.4){$z_2=x_2+iy_2$}\rput[l](0.2,2.3){$z_1=x_1+iy_1$}\rput(0.5,4.){\Large${\mathtt U}$}
    \psarc{->}(3.5,1){0.8}{0}{80}
    \psarc{->}(3.5,1){0.7}{0}{148}
    \rput(4.5,1.5){$\theta_2$}\rput(3.2,1.9){$\theta_1$}
    \rput(0.2,1.2){$-1$}\psdot(0.5,1)
    \rput(6.7,1.2){$1$}\psdot(6.5,1)
  \end{pspicture}
  \caption{Left: Strip with domain-walls at $w=0$ and $w=i2\tz$ and points at
    $r_i+i\t$, with $\t$ that must be analytically continued to $\tz+it$.
Right: Corresponding geometry on the upper half-plane after the mapping
(\ref{map}).
For the two-point correlation functions,
the Euclidean distance between the two points is given by
$\rho^2=(x_1-x_2)^2+(y_1-y_2)^2$.
}
  \label{fig:uhp}
\end{figure}
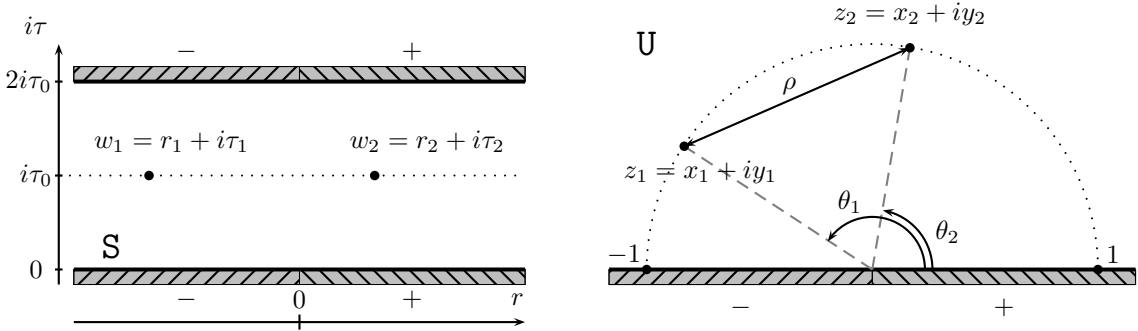

To obtain the time-dependence of a correlation function evolving from a state with one domain-wall (let say at
$r=0$ for simplicity), we need to analytically continue the same correlation function calculated in a strip of
width $2\t_0$ with boundary condition changing operators at $r=0$ on both the borders of the strip. The strip
complex coordinate is $w=r+i\tau$ with $0<\tau<2\tz$. The operators needed for the correlation functions are
then inserted at $\t=\tz+i t$. This geometry is depicted on the left of figure \ref{fig:uhp}. Note that we have
set the sound velocity to unity.

The strip geometry can be obtained by the conformal mapping of the
upper half-plane $z=x+iy$ with a domain-wall in the origin (and
one at $\infty$)
\be
z= \frac{e^{\pi w /(2 \tz)}-1 }{e^{\pi w /(2 \tz)} + 1} =
\tanh\left(\frac{\pi w}{4\tz}\right)\,.
\label{map} \ee 
We indicate in all the paper the upper half-plane with ${\mathtt U}$ and the 
strip with ${\mathtt S}$. $\theta$ is the argument of $z$ in ${\mathtt U}$, as
shown in figure \ref{fig:uhp} (right panel). Note that the conformal map used
here is different from the one used in Refs. \cite{cc-06,cc-07} 
$w(z)=(2 \tau_0/\pi) \ln z$ since we fix two marked
points on the boundary (the two domain-walls).

In the case where $\cal O$ is a product of local \em primary \em scalar 
operators $\Phi_i(w_i,\bar w_i)$, the expectation value in the strip is
related to the one in ${\mathtt U}$ by the standard transformation 
\be
\langle\prod_i\Phi_i(w_i,\bar w_i)\rangle_{\mathtt S} 
=\prod_i|w'(z_i)|^{-\Delta_i}
\langle\prod_i\Phi_i(z_i,\bar z_i)\rangle_{\mathtt U}\,, 
\label{mapO} 
\ee 
where $\Delta_i=2 h_i$ is the bulk scaling dimension of $\Phi_i$. Often in the
following we use the shorthand $\Phi_i(z)$ for $\Phi_i(z,\bar z)$. Note that,
as usual, when comparing to lattice models, an eventual expectation value of
$\Phi_i$ in the ground state of $H$ is supposed to have been subtracted
off. The asymptotic real time dependence is obtained via the analytic
continuation $\tau\to\tz+it$, and taking the limit $t,r_{ij}\gg \tz$. 

In the following sections we apply this method to several specific cases.

\section{The Ising model}
\label{sec3}

We start our analysis from the Ising model for which several correlation
functions (of the two primary operators spin and energy)
in the upper half-plane with a domain-wall have been already
calculated by Burkhardt and Xue \cite{bx-90} (using specific results for four-
and six-point correlation functions previously derived \cite{6p}).
In the beginning, we shall focus our interest to domain-walls created by 
changing fixed boundary conditions from $a=+$ to $b=-$.
We refer to this case as ``$+-$ initial conditions''. The initial position of 
the domain wall is at $r=0$ (unless noted otherwise) and $a$ on its positive 
side. Then we separately discuss the case $a=+$ and $b=\mbox{free}$, a 
domain-wall defined via changing from fixed to free boundary conditions. 
We will abbreviate this case by ``$+f$
initial conditions''. All results here should be relevant for the time 
evolution of e.g. the Ising chain in
a transverse field with Hamiltonian 
$H=- \frac{1}{2} \sum_i [s^x_i s^x_{i+1}+h s_i^z]$ (note that the state 
$|+\rangle$ here corresponds a spin aligned in the $x$ direction). 
This model is easily diagonalized through a Jordan-Wigner transformation 
followed by a Bogoliubov rotation, that map it onto a free fermion lattice 
model (see e.g. appendix of Ref. \cite{cc-05}). 
It is critical at $h=1$ and hence described there by a CFT with central charge
$c=1/2$.

\subsection{$+-$ initial condition: one-point functions}

\paragraph{Spin.}
The $\sigma$ expectation value with $+-$ boundary condition in ${\mathtt U}$
is \cite{bx-90} 
\be
\langle\sigma\rangle_{+-}^{\mathtt U}= {\cal A} \frac{\cos\theta}{y^{1/8}}
\,, \label{sigIs}
\ee 
where ${\cal A}$ is a non-universal amplitude, that can be fixed in a universal way through the ratio
with the amplitude of the bulk two-point function \cite{c-86,cl-91} (but we will never use the explicit value of
${\cal A}$). This correlation function satisfies the correct (bulk) limiting behavior as $y\to +\infty$. Hence
it is straightforward to read off that the scaling dimension of $\sigma$ is $\Delta_\sigma=2 h_\sigma=1/8$.

\begin{figure}[htpb]
  \centering
  \includegraphics[width=0.7\textwidth]{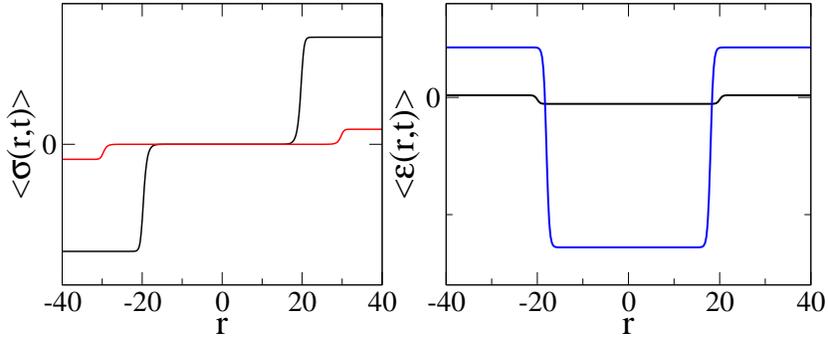}
  \caption{Space-time dependence of one-point correlation functions
   from a $+-$ initial condition.
   Left: $\langle\sigma(r,t)\rangle$ for $t=20,30$ as function of $r$.
   Right $\langle\varepsilon(r,t)\rangle$ for $t=18,20$ as function of $r$.
   Both the $y$ axes are in arbitrary units and we fixed $\tz=1$.
}
  \label{fig:eplusminus}
\end{figure}

The magnetization profile in the strip with the central domain-wall is
obtained by using Eq. (\ref{mapO}) and the conformal mapping (\ref{map}) 
\be 
\langle\sigma\rangle_{+-}^{\mathtt S}={\cal A} \sqrt2
\left(\frac\pi{2\tz}\right)^{1/8} \frac{\sinh (\pi r/2\t_0)}{(\cosh(\pi r / \t_0) -\cos(\pi \t/ \t_0))^{1/2}}
\frac1{[\sin(\pi\t/2\t_0)]^{1/8}}\,. 
\ee 
To extract the real-time evolution we analytically continue to
$\tau=\tau_0+i t$, obtaining ($r$ stands for the distance from the
domain-wall)
\be 
\langle
\sigma(r,t)\rangle={\cal A} \sqrt2 \left(\frac\pi{2\tz}\right)^{1/8}
\frac{\sinh (\pi r/2\tz)}{(\cosh(\pi r /\tz) 
+\cosh(\pi t/ \tz))^{1/2}} \frac1{[\cosh(\pi t/2\t_0)]^{1/8}}\,, 
\label{spm} 
\ee 
which for asymptotic large time and distances $\tau_0\ll r,t$ reads 
\be 
\langle \sigma(r,t)\rangle = 
{\cal A} \left(\frac{\pi}{\tau_0}\right)^{1/8}{\rm sign}(r)\times
\cases{ e^{-\pi t/16\tau_0}& $t< |r|$\,, \cr 
e^{-\pi t/16\tau_0} e^{-\pi(t-|r|)/2\tau_0} & $t> |r|$\,. } 
\label{spmasy} 
\ee  
This result is depicted in the left panel of figure \ref{fig:eplusminus}.

Let us now discuss and interpret this first result.
It is the consequence of two competing effects:
\begin{itemize}
\item The states $+$ and $-$ are {\it not} eigenstates of the conformal
Hamiltonian.
Thus a half-line with $+$ (or $-$) initial condition evolves
according to the results of Refs. \cite{cc-06,cc-07}. In particular, we expect
$\langle\sigma(r,t)\rangle\propto e^{-\pi t/16\tz}$, that is nicely confirmed
by Eq. (\ref{spmasy}) in the case $t<|r|$.
Far enough from the origin, the system does not feel the domain-wall
in the initial state and the exponential decay of the one-point
function is the same as for a translational invariant initial state.
\item The effect of the domain-wall (encoded in the first term of
  Eq. (\ref{spm})) arrives at the point $r$ only at time $t=|r|$ (we recall
  that the speed of sound has been set to 1) and it is responsible for
  an additional exponential time decay $e^{\pi (t-|r|)/2\tz}$.
  This can be seen as a result of the quantum averaging of spins with opposite
  orientations.
\end{itemize}
As usual \cite{cc-06,cc-07}, the effect of finite $\tz$ is only to smooth
the time evolution close to the crossover point $t=|r|$.

\paragraph{Energy density.}
The time evolution of the energy-density operator can be obtained in analogous
way. In the upper half-plane we have \cite{bx-90} 
\be 
\langle\e \rangle_{+-}^{\mathtt U}={\cal B} \frac{4\cos^2\theta-3}{y}\,, 
\label{eIs}
\ee 
i.e.
the scaling dimension of $\e$ is $\Delta_{\epsilon}=2 h_\epsilon=1$ (as well known for Ising) and there is a
different non-universal constant ${\cal B}$. Conformal mapping to the strip and the analytic continuation is
worked out using the formulas in \ref{appA}. For $t,r\gg\tz$, we obtain
\be
 \langle \varepsilon(r,t)\rangle\stackrel{t,r\gg\tz}{=}
{\cal B}\left(\frac{\pi}{\tau_0}\right)e^{-\pi t/2\tau_0}\times
\cases{
    1& $t<|r|$,\cr
    -3
& $t>|r|$. } \ee See figure \ref{fig:eplusminus} (right) for illustration. The
interpretation of the result is similar to the previous case, except that,
contrary to the spin operator, the energy density of the $+$ and $-$ state are
equal and so the result is symmetric with respect to the axis $r=0$. 
For $t<|r|$, the time evolution is the same as in the absence of the
domain-wall that only plays a role for $t>|r|$. However, for the energy, the
domain-wall has a drastic influence, because the averaging of spins with
opposite direction changes abruptly (at least for $\tz\ll t,|r|$) the value of
the energy density. 

A cartoon of the time evolution of the one-point functions, e.g. in the 
transverse field Ising chain, goes as follows. 
The initial domain-wall state undergoes a spin-flip process which locally 
looks like $|-+\rangle \to (|++\rangle-|--\rangle)/\sqrt2$. 
This process does not affect the two-site 
magnetization (in the $x$ direction), but makes vanishing the single-site one,
that before was maximum. Also the energetic balance is positive since the 
singlet state has lower energy than the product one in the critical 
Ising chain.
Similar processes at later times $t>|r|$ are responsible for the negative 
local $\epsilon$ and the suppressed magnetization (as in Eq. (\ref{spmasy})
inside the causal-cone of the domain-wall. 
We stress that this is only a cartoon of what happens {\it locally} because 
the real-time evolved state is highly entangled on a global scale (see the 
section about entanglement entropy).

\subsection{+- initial condition: two-point functions}

\begin{figure}[htpb]
\centering
\includegraphics[width=0.7\textwidth]{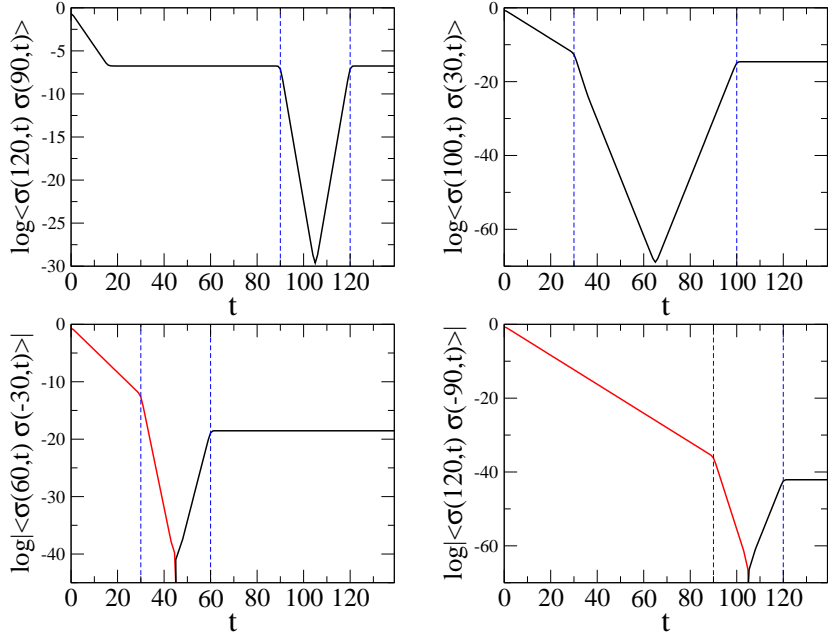}
\caption{Two-point function of the spin operator for four different quenches
with $(r_2,r_1)=(120,90)$, $(100,30)$, $(120,-90)$, $(30,-60))$.
They are the exact results with $\tau_0=1$.
When a single $r_j$ is negative ($r_j<0$) we plot the absolute value.
Signs are indicated by the colors of the curves:
black corresponds to $+$ and red to $-$.}
\label{regimes}
\end{figure}

\paragraph{Spin-spin.}
Let us start with the spin-spin correlation function. In the upper half-plane $\mathtt U$, we abbreviate
$\sigma(z_k)$ with $z_k=x_k+iy_k$ by $\sigma(k)$, $k=1,2$.
According to Ref. \cite{bx-90}, for boundary conditions $+-$, as
shown in figure \ref{fig:uhp}, the two-point function is given by
\bea\fl
\left\langle\sigma(1)\sigma(2)\right\rangle_{+-}^{\mathtt U}&=&
\frac{{\cal A}^2}{\sqrt{2}}\frac{1}{(y_1y_2)^{1/8}}\left(\sqrt{u+u^{-1}}\cos\theta_1\cos\theta_2
+\frac{u-u^{-1}}{\sqrt{u+u^{-1}}}\sin\theta_1\sin\theta_2
\right)\nonumber\\
&=&\frac{\left\langle\sigma(1)\right\rangle_{+-}^{\mathtt U}
   \left\langle\sigma(2)\right\rangle_{+-}^{\mathtt U}}{\sqrt{2}}
\sqrt{u+u^{-1}}\left(1+\frac{u^2-1}{u^2+1}\tan\theta_1\tan\theta_2\right),
\eea
with the same amplitude ${\cal A}$ as for the one-point function.
Here $\theta_i$ are polar angles as indicated in figure \ref{fig:uhp}, and $u$ an auxiliary variable defined via
\begin{equation}
  u=\left(1+\frac{4y_1y_2}{\rho^2}\right)^{1/4}.
  \label{eqn:defu}
\end{equation}
Always using Eqs. (\ref{mapO}) and (\ref{map}) we obtain the strip result
${\mathtt S}$
\be\fl
 \left\langle\sigma(1)\sigma(2)\right\rangle_{+-}^{\mathtt S}=
\frac{\left\langle\sigma(1)\right\rangle_{+-}^{\mathtt S} \left\langle\sigma(2)\right\rangle_{+-}^{\mathtt
S}}{\sqrt{2}}{\sqrt{u+u^{-1}}} \left(1+\frac{u^2-1}{u^2+1}\,\frac{\sin(\pi \t_1/2\tau_0)\sin(\pi
\t_2/2\tz)}{\sinh(\pi r_1/2\tz)\sinh(\pi r_2/2\tz)}\right), \ee where the Jacobians from conformal mapping have
canceled out. The first factor does not contain any substantially new information since it is the product of two
one-point functions. Analytic continuation of the various terms is performed
in \ref{appA}. 
Putting everything together, we can write the time evolution of the two-point 
correlation function for $t,r_i,|r_1-r_2|\gg \tz$ as 
\be\fl
\left\langle\sigma(1)\sigma(2)\right\rangle= \cases{ \langle\sigma(1)\rangle \langle\sigma(2)\rangle &
$2t<|r_1-r_2|$,\cr \frac{ \langle\sigma(1)\rangle \langle\sigma(2)\rangle}{\sqrt2} e^{ \pi
(2t-|r_1-r_2|)/16\tau_0}\left(1+ s_1 s_2 e^{ \pi (2t-(|r_1|+|r_2|))/2\tau_0}\right) & $2t>|r_1-r_2|$, } \ee with
$s_i={\rm sign}\, r_i$, where $u \approx 1$ in the first regime and $u \to \infty$ in the second (since
$2 t < |r_1-r_2| \leq |r_1|+|r_2|$ one can check from (\ref{eqn:u4continued}) 
that there is no third regime where small $u-1$ terms could mix with a large 
contribution from the $s_1 s_2$ term). 
This divides in different sub-regimes according to the relative values of 
$r_1,r_2, |r_1-r_2|/2,t$. More precisely, we can distinguish two
main regimes:
\begin{itemize}

\item $2 t <|r_1-r_2|$: the two points evolve independently since the horizon effect prevents from mutual interaction.
The correlation decays as the product of the two one point functions and the connected correlation can be said
to vanish (more precisely, its decay is subdominant). There are two possible sub-regimes: \be
\fl\left\langle\sigma(1)\sigma(2)\right\rangle = {\cal A}^2 \left(\frac\pi{\tz}\right)^{1/4} s_1 s_2 e^{- \pi t /8\tau_0} \times \cases{
1 & $t<\min(|r_1|,|r_2|)$\,,\\
e^{\pi (t-|r_2|)/2\tau_0}  & $t>|r_2|$\,. } \ee the second regime exists only 
if $|r_1-r_2|>2\min(|r_1|,|r_2|)$ and we have written it in the case 
$|r_2|<|r_1|$ (we will adopt this convention in this
subsection). Let us denote Ia and Ib these two subregimes.

\item $2t>|r_1-r_2|$: One has then the following possible sub-regimes (which we can denote IIa,b and c, respectively):

\begin{itemize}

\item For $|r_1-r_2|/2<t<|r_1|,|r_2|$ we find
\be \left\langle\sigma(1)\sigma(2)\right\rangle = \frac{{\cal A}^2}{\sqrt{2}} \left(\frac\pi{\tz}\right)^{1/4} e^{- \pi |r_1-r_2|/16\tau_0}\,. \ee This is a
quasi-equilibrium regime. Both points have space-like separation to the domain-wall, but time-like to each
other, hence they equilibrate as if no domain-wall was present. Obviously this regime exists only if the
two points are on the same side of the domain-wall $s_1 s_2=1$. It only exists for a finite time interval, as
eventually the domain-wall influence will be felt.

\item For $|r_2|<t< |r_1|$ we have
\be \fl\left\langle\sigma(1)\sigma(2)\right\rangle = \frac{{\cal A}^2}{\sqrt{2}}  \left(\frac\pi{\tz}\right)^{1/4}  e^{- \pi |r_1-r_2|/16\tau_0} \times \cases{
s_1 s_2 e^{-\pi (t-|r_2|)/2\tau_0} & $2t<|r_1|+|r_2|$ \,,\\
e^{\pi (t-|r_1|)/2\tau_0}  & $2t>|r_1|+|r_2|$\,, } 
 \ee 
with an intermediate regime when correlation again evolves in time. 
Its absolute value first decreases down to the minimal value
$|\left\langle\sigma(1)\sigma(2)\right\rangle|_{\rm min} 
\propto e^{- \pi (|r_1-r_2| + 4 (|r_1|-|r_2|)) /16\tau_0}$
reached at time $t=(|r_1|+|r_2|)/2$. 
After that time it increases again. In addition, if the points are on
opposite side of the domain-wall, there is a sudden change in sign (i.e. on 
scale $\tau_0$) of the correlation
at time $t=(|r_1|+|r_2|)/2=|r_1-r_2|/2$ hence only the second (increasing) 
part of this subregime exists. Note that although the absolute value vanishes 
on time scales smaller than $\tau_0$, it is possible to define a 
minimum value on scales much larger than $\tau_0$, as can be seen on the two
bottom figures in Fig. \ref{regimes}.

\item Very large times $2t> |r_1-r_2|, 2|r_1|, 2|r_2|$, almost all terms
cancel out and we may simplify to \be \left\langle\sigma(1)\sigma(2)\right\rangle = \frac{{\cal A}^2}{\sqrt{2}} \left(\frac\pi{\tz}\right)^{1/4}  e^{- \pi
|r_1-r_2|/16\tau_0}\,, \ee i.e. the correlation function is the same as in a thermal state with an effective
temperature $\beta_{\rm eff}=4 \tau_0$ as in the case without the domain-wall:
the system has ``relaxed'' it.

\end{itemize}
\end{itemize}

Let us discuss these findings critically and describe which of these regimes are encountered as the system
evolves. We start with the case of $r_1$ and $r_2$ on the same side of the domain-wall, i.e. without loss of
generality $r_1>r_2>0$. Then $r_1-r_2>0$ can either be larger or smaller than $2 r_2$. If $(r_1-r_2)/2<r_2$ then
the evolution can be summarized as follow:
\begin{itemize}
\item $2t<r_1-r_2$: the connected two-point function is constantly zero.
\item $(r_1-r_2)/2<t<r_2$: the two-point function has a plateau at
  $e^{-\pi (r_1-r_2)/16\tau_0}$. Until this time the two points at $r_1,r_2$
  did not yet feel the presence of the domain-wall.
  In fact the result is exactly the evolution found from a
  boundary state without domain-wall \cite{cc-06,cc-07}, in analogy with the
  one-point functions.
\item At $t=r_2$, the part of the system under consideration finally enters
  in the horizon of the domain-wall and the two-point function
  starts evolving again with an exponential decreasing up to the time
  $(r_1+r_2)/2$. At this time, the correlation starts increasing toward the
  asymptotic value.
\item At $t=r_1$ the system reaches its asymptotic value and the two point correlation does not evolve
  anymore. The domain-wall has been relaxed.

\end{itemize}
An example for $(r_1-r_2)/2<r_2$ is reported on the top left figure \ref{regimes}, where $(r_1,r_2)=(120,90)$.
The succession of regimes is thus Ia, IIa, IIb, IIc.

The case $2r_2<r_1-r_2$ turns out to be slightly different, and we give an example with $(r_1,r_2)=(100,30)$ in
the top right figure \ref{regimes}. In this case the sub-system feels the domain-wall before it is quasi-relaxed
and so the succession of regimes is now Ia, Ib, IIb and IIc.

When the two points are on different sides of the domain-wall, clearly $|r_1-r_2|/2$ (being the average of their
absolute value) is always in between  $|r_1|$ and $|r_2|$, so that the plateau (quasi-equilibrium) regime IIa
never exists. Hence the succession of regimes is Ia, Ib, IIb (second half only) and IIc. Let us point out an
interesting difference to the previous cases: for short time the two-point correlator is negative, since $s_1
s_2=-1$, and changes sign at $t=|r_1-r_2|/2$ (when the systems goes from regime Ib to IIb) in order to reach a
positive asymptotic value. This can be easily understood: in fact for $2t<|r_1-r_2|$ each point is out of the
horizon of the other and evolves keeping its initial sign. Connected correlations start forming only after
$t=|r_1-r_2|/2$ when the two-point function changes sign.

\begin{figure}[htpb]
\centering
  \includegraphics[width=0.7\textwidth]{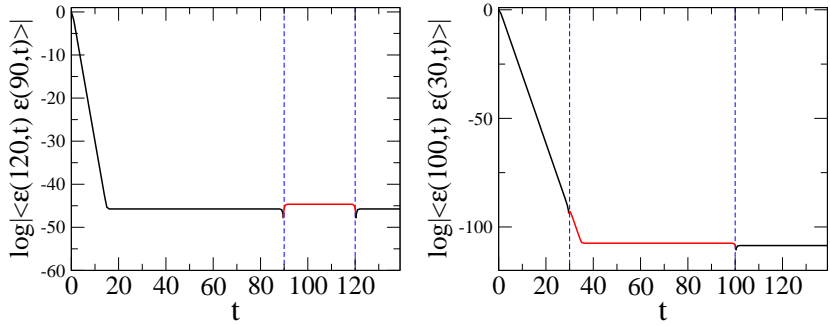}
  \caption{Absolute value of the energy-energy two-point function
  $\langle\epsilon(r_1,t)\epsilon(r_2,t) \rangle$ as a function of $t$ for
  $(r_1,r_2)=(120,90)$, and $(100,30)$. We fix $\tau_0=1$.
   Signs are indicated by the colors of
   the curves: black corresponds to $+$ and red to $-$.}
  \label{fig:eepm}
\end{figure}

As a general conclusion, although the correlation reaches the same asymptotic large time equilibrium value as if
the domain-wall was absent, we found that it does exhibit an interesting non-monotonic time dependence, which shows that the domain-wall has a disordering effect on the system at intermediate times. A complementary picture of the various regimes is given below (see Fig. \ref{fig:cones}).

\paragraph{Energy-energy.}
We proceed to the two-point function of the energy-density operator 
$\varepsilon$ which has been calculated in Ref. \cite{bx-90} in the upper 
half-plane. We now map it to the strip and rewrite the resulting correlation 
function in a suitable way
\bea 
\fl \langle \e(1)\e(2)\rangle_{+-}^{\mathtt U}= 
\langle \e(1)\rangle_{+-}^{\mathtt U}\langle\e(2)\rangle_{+-}^{\mathtt U} 
\nonumber\\ \fl \qquad\qquad\times
\left[1+(u^4-1)\frac{1-4\sin^2(\theta_1-\theta_2)}{(1-4\sin^2\theta_1)(1-4\sin^2\theta_2)}
-\frac{u^4-1}{u^4}\frac{1-4\sin^2(\theta_1+\theta_2)}{(1-4\sin^2\theta_1)(1-4\sin^2\theta_2)} \right]\,.
\label{sumen} 
\eea 
Transport to the strip is simple and the analytical continuation of the
various terms is presented in \ref{appA}. Note that the factor $u^4-1$ 
indicates that the connected two-point function vanishes for $2t<|r_1-r_2|$, 
as it can be seen from Eq. (\ref{eqn:u4continued}). 
As above, it corresponds to the independent evolutions of the two points, 
which are not yet causally connected. Hence we focus on $2t>|r_1-r_2|$, when 
the two points interact. 
In this regime $u^4 \to \infty$ from Eq. (\ref{eqn:u4continued}) and one finds 
that only the second term in the parenthesis of Eq. (\ref{sumen}) contributes.
By analogy with the spin-spin correlation function, we find three different 
sub-regimes for $2t>|r_1-r_2|$:
\begin{enumerate}
\item For $t>|r_1|,|r_2|$, the system displays thermal like correlations 
and, in fact, we find
\be 
\langle \varepsilon(1)\varepsilon(2)\rangle_{+-}= \mathcal{B}^2\left(\frac{\pi}{\tau_0}\right)^2\,
e^{-\pi|r_1-r_2|/{2\tau_0}}\,, 
\label{eeasym} 
\ee 
with the correct prefactor, and with $\beta_{\rm eff}= 4\tau_0$ as
it should be from general arguments (see below).
\item In the opposite limit $t<|r_1|,|r_2|$ (always with $2t>|r_1-r_2|$),
  the correlation function is given by Eq. (\ref{eeasym}), with the 
  {\it same} prefactor. The two points are quasi-equilibrated since they do
  not yet feel the presence of the domain-wall.
\item  For $|r_2|<t<|r_1|$ (we assume without loss of generality $|r_1|>|r_2|$)
  we find
\begin{equation}
\langle \varepsilon(1)\varepsilon(2)\rangle_{+-}=
-3\mathcal{B}^2\left(\frac{\pi}{\tz}\right)^2\,e^{-\pi|r_1-r_2|/{2\tau_0}}\,,
\end{equation}
where the result is the same as in the other two regimes, only being
multiplied by $-3$.
At variance from the spin correlation function, does not depend on
time (apart from an obvious smoothing close to the changing times 
$t=|r_1|,|r_2|$). This apparently strange phenomenon, visible in 
Fig. \ref{fig:eepm}, can be traced back to the fact that 
$\langle\e(r,t)\rangle$ 
for $t>|r|$ does not have any further exponential suppression, 
but only a discontinuity at $|r|$.
\end{enumerate}

\paragraph{Spin-energy.} In the upper half-plane $\mathtt{U}$, the 
spin-energy correlation function reads
\be
\langle \sigma(1)\e(2)\rangle_{+-}^{\mathtt U}=
\langle \sigma(1)\rangle_{+-}^{\mathtt U}\langle \e(2)\rangle_{+-}^{\mathtt U}
\times\frac1{u^2}\left(\frac{u^4+1}{2}+ (u^4-1)
\frac{\tan\theta_1\sin 2\theta_2}{4\cos^2\theta_2-3}\right)\,.
\ee
The passage to the strip and the analytic continuation is easily done using
formulae in \ref{appA}.
Again for $2t<|r_1-r_2|$ where $u\sim 1$, this two-point function is just the
product of the one-point, in agreement with causality.

\begin{figure}[htpb]
  \centering
  \includegraphics[width=0.7\textwidth]{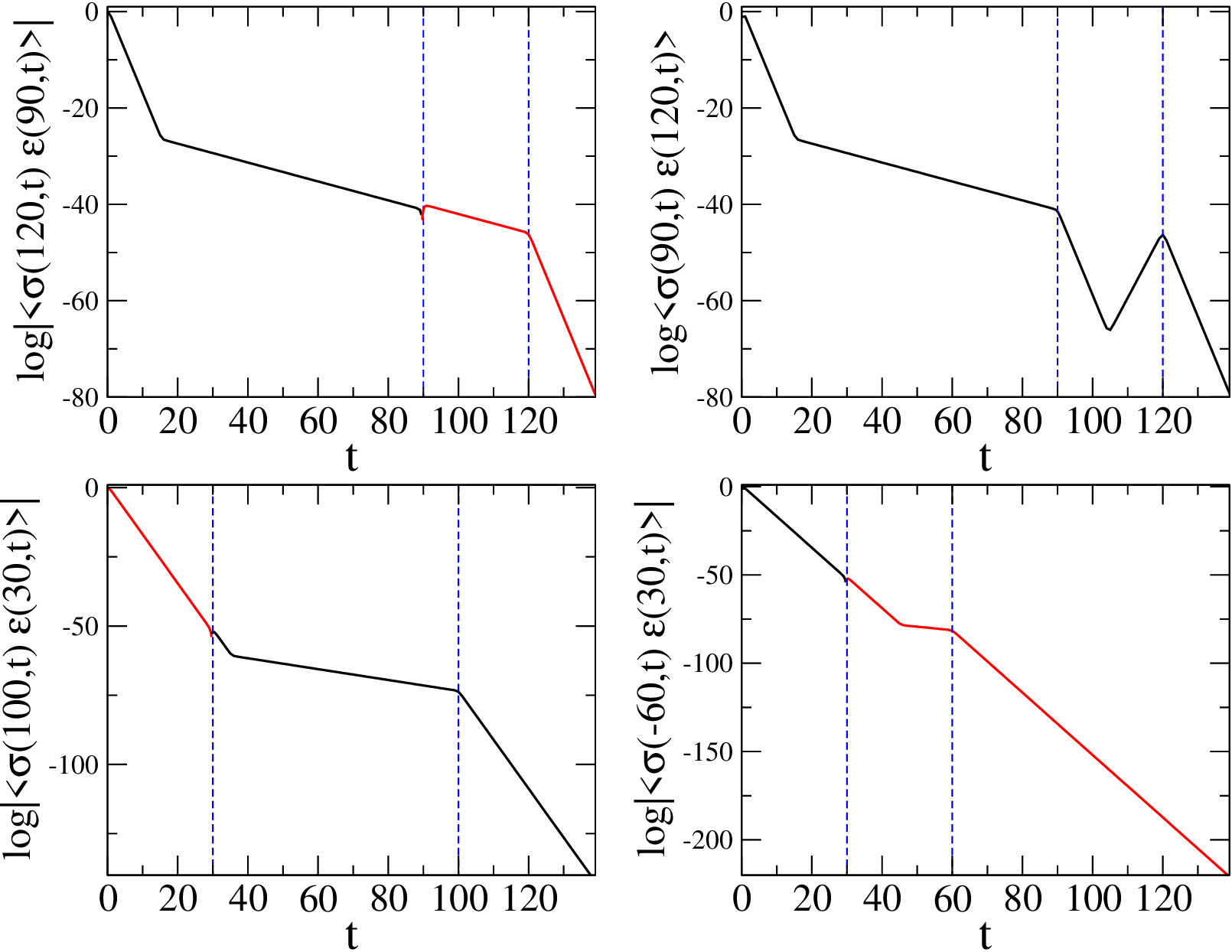}
  \caption{Absolute value of the spin-energy two-point functions
  $\langle\sigma(r_1,t)\varepsilon(r_2,t) \rangle$ as a function of $t$ for
  $(r_1,r_2)=(120,90)$, $(90,120)$, $(100,30)$, and $(-60,30)$.
  We fix $\tau_0=1$. Signs are indicated by the colors of
  the curves: black corresponds to $+$ and red to $-$.}
  \label{fig:espm}
\end{figure}

Slightly more involved is the opposite case $2t>|r_1-r_2|$, where
asymptotically we have 
\be 
\fl  \langle \sigma(1)\e(2)\rangle_{+-}= 
\langle \sigma(1)\rangle_{+-}\langle \e(2)\rangle_{+-}\,
e^{\pi(2t-|r_1-r_2|)/4\tz}\left(\frac12+\frac{2 s_1s_2 e^{\pi(t-|r_1|)/2\tz}
e^{-\pi|t-|r_2||/2\tz}}{1-4\Theta(t-|r_2|)}\right). 
\ee 
Here $\Theta(x)$ denotes the Heaviside function which is $1$ for $x>0$, and 
$0$ otherwise. We will not give a detailed discussion of the various
sub-regimes. Instead we give some examples in figure \ref{fig:espm} where we
illustrate the time evolution for several choices of $r_1$ and $r_2$. 
Amongst the various cases, the most interesting and new feature when compared 
to $\langle\sigma\sigma\rangle$ and $\langle\e\e\rangle$ is the very
large time behavior: as $t$ exceeds any other scale, i.e. 
for $t>|r_1|,|r_2|,|r_1-r_2|/2$,
we find that 
\be 
\langle \sigma(1)\e(2)\rangle_{+-} = 
2 {\cal A} {\cal B} s_2 e^{-\pi t (1/16+1/2) /\tz} e^{-(|r_1-r_2|- 2
|r_2|)/4\tz}, 
\ee 
which decays to zero for $t\to\infty$. Indeed this feature is in agreement
with the well-known fact that the finite-temperature equilibrium value of a 
two-point function of operators with different scaling dimensions vanishes 
(see also \ref{appB}).

\subsection{+f initial condition: one-point functions}

Let us consider different boundary conditions: we suppose a $+$ state on the
half line from $0$ to $+\infty$. On the negative real axis the spins are left 
free to take either value $+$ or $-$.

\begin{figure}[htpb]
\centering
\includegraphics[width=0.7\textwidth]{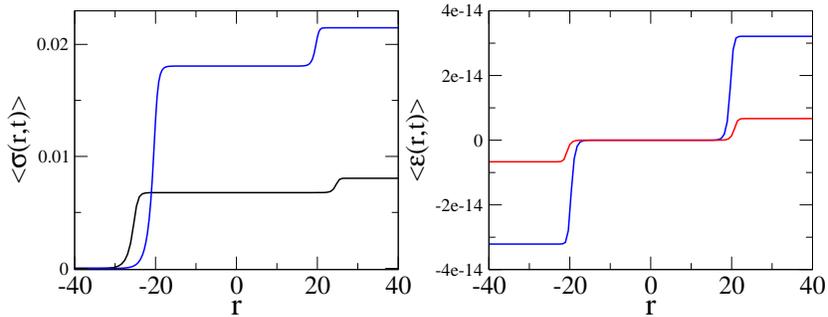}
\caption{Left: One-point function $\langle\sigma(r,t)\rangle$ with $+f$ 
initial conditions for $t=20$ (blue) and $t=25$ (black).
Right: One-point function $\langle \epsilon(r,t)\rangle$ for $t=20$ (blue) and
$t=21$ (red).
We fix $\tau_0=1$ and the $y$-axis are measured in arbitrary units.}
  \label{fig:plusfree}
\end{figure}

\paragraph{Spin.}
In the upper half-plane the one-point correlation function of the spin is
\cite{bx-90}
\begin{equation}
\left\langle\sigma\right\rangle_{+f}=
\mathcal{A}\,y^{-1/8}\sqrt{\cos(\theta/2)}.
\end{equation}
Using results from \ref{appA}, the one-point function is asymptotically
given by
\be
\langle\sigma(r,t)\rangle_{+f}=
  \mathcal{A}\left(\frac{\pi}{\tau_0}\right)^{1/8}
e^{-\pi t/16\tau_0}
  \cases{
    2^{-1/4}& $|r|<t$,\cr
    \Theta(r) &$|r|>t$.
  }
\ee 
See figure \ref{fig:plusfree} (left panel) for an illustration. Note that
the amplitude vanishes on the space like negative region $r<-t$ as for 
homogeneous free boundary conditions, and in fact
matches the result of Eq. (\ref{spmasy}) on the space-like positive side
$r>t$. Inside the time-like cone it takes a non trivial value.

\paragraph{Energy.}
In ${\mathtt U}$, the one-point function for the energy is \cite{bx-90}
\begin{equation}
\langle \e\rangle_{+f}=\frac{\cal B}{y}\,\cos\theta,
\end{equation}
that (a part the constant and the scaling dimension) is the same as the
spin expectation value with $+-$ conditions and so we have
the time evolution as
\be
 \langle \e(r,t)\rangle_{+f}
\simeq {\cal B}\left(\frac{\pi}{\tau_0}\right)e^{-\pi t/2\tau_0}
  {\rm sign}\,r
  \cases{
    e^{-\pi(t-|r|)/2\tau_0} & $|r|<t$,\cr
        1& $|r|>t$.
}
\ee
An illustration is given in figure \ref{fig:plusfree} (right panel).

\subsection{$+f$ initial condition: two-point functions}

\paragraph{Spin-spin.} Again, Ref. \cite{bx-90} gives the half-plane result
\bea\fl
 \left\langle\sigma(1)\sigma(2)\right\rangle_{+f}&=&
\frac{{\cal A}^2}{\sqrt{2}}\frac{1}{(y_1y_2)^{1/8}}
\sqrt{(u+u^{-1})\cos(\theta_1/2)\cos(\theta_2/2)
+(u-u^{-1})\sin(\theta_1/2)\sin(\theta_2/2)}\nonumber\\
&=&\frac{\left\langle\sigma(1)\right\rangle_{+f} \left\langle\sigma(2)\right\rangle_{+f}}{\sqrt{2}}\,
{\sqrt{u+u^{-1}}}\, \sqrt{1+\frac{u^2-1}{u^2+1}\tan(\theta_1/2)\tan(\theta_2/2)}. \label{bottom} \eea

\begin{figure}[htpb]
\centering
\includegraphics[width=0.7\textwidth]{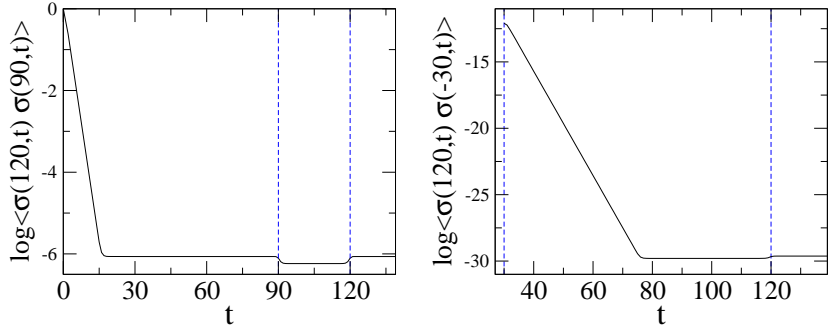}
\caption{
Spin-spin two-point functions
$\langle\sigma(r_1,t)\sigma(r_2,t) \rangle_{+f}$ as a function of $t$ for
$(r_1,r_2)=(120,90)$ (left) and $(120,-30)$ (right).
We fix $\tau_0=1$. When at least one $r$ is negative, the two-point function
is practically zero outside the two causal cones.
}
\label{fig:fp}
\end{figure}

Relevant formulae for conformal mapping to the strip and analytic continuation are given in \ref{appA}.
One finds similar regimes as for the $+-$ initial conditions except that 
this two-point function is always positive, as a consequence of the 
non-negativity of one-point correlators. 
The main features are now standard. 
For time such that $2 t < |r_1-r_2|$ the two-point correlation function is 
just the product of the one-point ones, as $u \to 1$ in Eq. (\ref{bottom}).
For larger times, it is important to note that in the limit 
$u\to\infty$ one should be careful in using the second line in 
Eq. (\ref{bottom}) because it has the form $0\times\infty$. 
It is simpler to use the first line properly mapped to the strip 
(as explained in \ref{appA}) that is always finite.
For times such that $2 t > |r_1-r_2|$ one finds, using that $u \to \infty$, 
a pseudo-thermal behavior with effective temperature $\beta_{\rm
  eff}=4\tau_0$, for all three possible subregimes (noted IIa,b,c above), with
\bea
\langle \sigma(r_1,t)\sigma(r_2,t)\rangle_{+f} = \frac{1}{\sqrt{2}} {\cal A}^2
\left(\frac{\pi}{\tau_0}\right)^{1/4} e^{-\pi|r_1-r_2|/16\tz}\,, \label{same}
\eea
for both small and large time regimes $t< |r_1|,|r_2|$ and $t>|r_1|,|r_2|$. 
More surprisingly one finds the same {\it time independent} 
result (\ref{same}) for the intermediate time regime  $|r_2| < t< |r_1|$, 
but with an amplitude multiplied by the factor $2^{-1/4}$. 
This is at variance with the behaviour of the $+-$ case, and it is illustrated 
by two examples in figure \ref{fig:fp}, where two different pairs of 
values of $r_1,r_2$ are considered. 
We conclude that this kind of mixed domain-wall can be
``dissipated'' by the system, too.

The other two-point correlation functions with $(+f)$ boundary condition 
are known in the half-plane from Ref. \cite{bx-90}. 
We do not report their analysis here, since the main features 
of the Ising universality class are already present in the correlation 
reported so far.
We rather highlight general features in the next section and show 
which information can be obtained from general scaling arguments only, 
without detailed knowledge of the full correlators.

\section{General CFT results}
\label{sec4}

Some of the results previously obtained for the Ising CFT with particular 
boundary conditions have a direct generalization to an arbitrary theory 
with any domain-wall in the initial state between two boundary conditions 
$a$ (to its right) and $b$ (to its left). 
In fact any $n$-point correlation $(\zeta_1-\zeta_2)^{-2 \Delta_{ab}} 
\langle \Phi_1(z_1) \cdots \Phi_n(z_n) \rangle_{ab}^{\mathtt U}$ 
of primary operators satisfies the same 
differential equation as the full plane $2n+2$-point correlation
$\langle \Phi_1(z_1) \Phi_1(\bar z_1) \cdots \Phi_n(z_n) \Phi_n(\bar z_n) \Phi_{ab}(\zeta_1) \Phi_{ab}(\zeta_2) \rangle$ and can
be constructed from the same solutions, as discussed in full details 
in Ref. \cite{bx-90}. 
$\zeta_1$ and $\zeta_2$ stand for the start and end positions of the interface 
on the real axis, taken here to be $(0,\infty)$, and $\Delta_{ab}$ is the 
scaling dimension of the $ab$ boundary changing operator $\Phi_{ab}$. 
Let us discuss general consequences of this property.

The one-point function of any primary operator in the upper half-plane
displays the scaling form
\be
\langle \Phi(z)\rangle_{ab}^{\mathtt U}={\cal A}_{ab} y^{-\Delta_\Phi}
f_{ab}^{\Phi}(\cos\theta)\,,  \label{onepointgen} 
\ee
where $\Delta_\Phi$ is the scaling dimension of $\Phi$ and all the dependence
on the boundary state is in $f_{ab}^{\Phi}(\cos\theta)$,
with, as usual, $\cos\theta=x/(x^2+y^2)^{1/2}$. 
In fact, $f_{ab}^{\Phi}(\cos\theta)$ explicitly depends on the boundary CFT 
and is related to hypergeometric functions
of the corresponding full plane $4$-point correlation.

A large class of primary operators displays a one-point function with a 
close analytic form (given in Ref. \cite{bx-90}), 
but in general they are quite complicated.
However, from the previous examples, it is clear that
we do not need the complete knowledge
$f_{ab}^{\Phi}(\cos\theta)$ to understand the quench.
Indeed upon mapping onto the strip and analytical continuation we already found
\bea 
\cos\theta&\to& {\rm sign}\, r \times 
\cases{ 1                    & $t<|r|$,\cr
        e^{-\pi(t-|r|)/2\tz} & $t>|r|$.
}\nonumber \\
&& \label{cosf} 
\eea
Hence we need only the behavior for $\cos\theta\sim 0,\pm 1$.
The case $\cos\theta\sim \pm1$ means that in the upper half-plane the point 
$z$ is far from the domain-wall and does not feel its effect.
Consequently, $f(\cos\theta)$ takes the same value as for homogeneous boundary 
conditions $a$ or $b$, related to $\cos\theta=1$ and $-1$ respectively, 
with associated amplitudes 
${\cal A}_{aa,bb}={\cal A}_{ab} f_{ab}^{\Phi}(\pm 1)$.
Conversely the case $\cos\theta=0$ regards any point inside the causal 
cone of  the domain-wall  (i.e. causally connected within the speed of sound). 
The amplitude $f_{ab}^{\Phi}(0)$ can be either zero or finite,
depending on the boundary CFT and the operator. 
We thus obtain 
\be
\langle \Phi(r,t)\rangle_{ab}=
{\cal A}_{ab} \left(\frac{\pi}{\tau_0}\right)^{\Delta_\Phi} 
e^{-\pi t \Delta_\Phi/2 \tz}\times
\cases{
f_{ab}^{\Phi}(\pm1) & $t<|r|$,\cr
f_{ab}^{\Phi}(0)    & $t>|r|$,
}
\label{1ptgen}
\ee
when $f_{ab}^{\Phi}(0) \neq 0$. 
Instead if $f_{ab}^{\Phi}(0)= 0$, the decay becomes faster, as 
$\sim e^{-\pi t \Delta_\Phi/2 \tz- \mu (\pi(t-|r|)/2\tz}$ for $t>|r|$ if 
$f_{ab}^{\Phi}(\cos\theta) \sim \cos^\mu\theta$ 
around $\cos\theta=0$, as in the case of the spin
correlation with $+-$ boundary condition for the Ising model (with $\mu=1$). 
An even more detailed discussion of the one-point functions is reported in 
Sec. \ref{sec6} in connection with SLE. In particular we show explictly 
in which cases $f_{ab}^{\Phi}(0)$ can vanish and that 
$f_{ab}^{\Phi}(\cos\theta)$ is analytic in zero, so in general $\mu=1$.

The (relative) simplicity of these one-point function is related to the
fact that bulk four-point functions can always be written as a general scaling
part times a function of the unique cross-ratio of the four points.
This is not the case for the two-point functions, which correspond to full 
plane six-point functions involving three independent cross ratios and the 
analysis is more complicated.
However, we can give a general characterization.
Considering a primary field $\Phi$ such
that $\langle\Phi(z)\rangle\neq0$, a scaling argument shows that
we can always write the two-point function in
${\mathtt U}$ with boundary condition $ab$ as
\be
\langle \Phi(z_1)\Phi(z_2)\rangle_{ab}^{\mathtt U}=
\langle \Phi(z_1)\rangle_{ab}^{\mathtt U}
\langle \Phi(z_1)\rangle_{ab}^{\mathtt U}
F_{ab}^{\Phi}(u,\cos\theta_1,\cos\theta_2)\,,
\label{2ptgen}
\ee
where the function $F_{ab}^\Phi$ is
(to the best of our knowledge) unknown in the general case, and related to
full-plane six-point functions. 
The variable $u$ is defined in Eq. (\ref{eqn:defu})
is (one of the possible) cross ratio of $z_{1,2}$ and $\bar z_{1,2}$ 
(note that comparing with Ref. \cite{cc-07} one has $\eta \to 1$ for 
$u \to \infty$ and $\eta \to 0$ for $u\to 1$). 
Formula (\ref{2ptgen})
still holds after mapping to the strip (replacing the index ${\mathtt U}$ by ${\mathtt S}$) with the appropriate expressions for the variables 
$u$ and $\cos \theta_i$ repeatedly used above and in \ref{appA}.
We thus find that the interesting regimes for the time evolution are 
encoded in particular limits of $F_{ab}^\Phi$. Namely:
\begin{itemize}
\item \textit{Short times $t<|r_1|,|r_2|$} and points on the same side of the 
initial domain-wall.
In this case, depending on the signs of $r_i$, we have $\cos\theta_i=\pm 1$. 
This corresponds to two points extremely close to one boundary condition, i.e.
\be
F_{ab}^{\Phi}(u,1,1)=F_{aa}(u)\,, \qquad F_{ab}^{\Phi}(u,-1,-1)=F_{bb}(u)\,,
\ee
where $F_{aa}(u)$ ($F_{bb}(u)$) corresponds to the two-point 
scaling function with homogeneous $a$ ($b$) boundary condition.
In this case the problem reduces to the one evolving from homogeneous
initial condition. 
Note that ${\cal A}^2_{aa} F_{aa}(u)= (u^4/4)^{\Delta_\Phi}$ for
$u \to \infty$ as one recovers the bulk behavior and, upon mapping to the 
strip, thermal like behavior in the subcase $t>|r_1-r_2|/2$ (see also below).

\item \textit{Short times $2t<|r_1-r_2|$, even on different sides.}
In this case $u\sim1$, i.e. the distance of the two points is much larger than
the distance from the boundary and
$F_{ab}^{\Phi}(1,\cos\theta_1,\cos\theta_2)=1$, meaning that the correlation
function evolves like the product of the one-point ones.
\item \textit{Larger times $2 t>|r_1-r_2|$}, the points are causally connected
and $u= e^{\pi(2t-|r_1-r_2|)/8\tz}\gg 1$. 
Scaling requires that 
$F_{ab}^{\Phi}(u,\eta_1,\eta_2) =u^{4 \Delta_\Phi}
\phi^\Phi_{ab}(\eta_1,\eta_2)$ 
in this large $u$ limit, so that the time dependence factorizes in front and 
\be 
\fl
\langle \Phi(r_1,t) \Phi(r_2,t)\rangle_{ab}=
{\cal A}_{ab}^2 \left(\frac{\pi}{\tau_0}\right)^{2 \Delta_\Phi} 
e^{- \pi \Delta_\Phi |r_1-r_2|/(2 \tau_0)} 
f_{ab}^{\Phi}(\eta_1) f_{ab}^{\Phi}(\eta_2) 
\phi^\Phi_{ab}(\eta_1,\eta_2). \label{last} 
\ee
The very large time subregime $t > |r_1|,|r_2|$ corresponds to 
$\eta_i=\cos\theta_i=0$. 
Consequently, in $\mathtt U$ the two points are deep in the
bulk and give the bulk two-point function, leading to the
thermal-like behavior. 
The other sub-regimes, when $t$ is in between the various ``critical''
times $|r_i|$ can be extracted from (\ref{last}) by considering the limits
$\cos\theta_i \to \pm 1,0$ and may lead to additional time dependence. 
Note that 
${\cal A}_{ab}^2 f_{ab}^{\Phi}(\eta_1) f_{ab}^{\Phi}(\eta_2) \phi^\Phi_{ab}(\eta_1,\eta_2)
\to 4^{-\Delta_\Phi}$ as $(\eta_1,\eta_2) \to (1,1)$ or $(-1,-1)$. 
\end{itemize}

The same reasoning can be applied to any $n$-point correlation function: for
short times all the points evolve incoherently, whereas for very long times
the correlation function is the same as in a thermal state at the effective
temperature $\beta_{\rm eff}=4\tz$. As for the homogeneous quench \cite{cc-07}
it is easy to understand the technical reason that gives the effective
temperature $4\tau_0$ both for asymptotic large times and also in the
intermediate regime. As usual, finite temperature correlations can be
calculated by studying the field theory on a cylinder of circumference $\beta
= 1/T$. In CFT a cylinder is usually obtained by mapping the complex plane
with the logarithmic transformation $w=\beta/(2\pi) \log z$, but one can
equivalently use the inverse of Eq. (2) $w=\beta/\pi {\rm arctanh} z$ 
that differs from the logarithm for a mapping of the real axis. 
The form for the two-point function of a primary operator in the strip 
depends in general on the function $F(u,\cos\theta_i)$, but
when we analytically continue, we find that $u\to\infty$, i.e. the 
original points in ${\mathtt U}$ effectively are far from the boundary 
(more precisely their relative distance is much less than the distance from
the real axis). 
Thus we get the same result as we would get
if we conformally transformed from the full plane to a cylinder, and from the
comparison of the two transformation the effective temperature is $\beta_{\rm
  eff }= 4\tau_0$. 
The same argument is easily worked out for the multi-point functions.

\section{The entanglement entropy}
\label{sec5}

Entanglement is a central concept in quantum information science. 
Moreover it is becoming a common tool to study and analyze extended quantum
systems because of its ability in detecting the scaling behavior close to
a quantum critical point \cite{ent-rev}.
A powerful measure is the block entanglement entropy
defined as follows. If the system is in a pure quantum state $|\Psi\rangle$,
$\rho=|\Psi\rangle\langle\Psi|$ is the density matrix.
Indicating with $A$ a spatial subset of the system (such as a
finite subset of spins in a spin chain) and with $\rho_A=\Tr_B \rho$ the reduced density
matrix of the subset $A$,
the entanglement entropy is just the corresponding Von Neumann entropy
\be
S_A=-\Tr_A \rho_A \ln \rho_A\,,
\ee
and analogously for $S_B$.
When $\rho$ corresponds to a pure quantum state, $S_A$ gives a measure of
the amount of quantum correlation between $A$ and $B$, and one can prove that $S_A=S_B$.

$S_A$ can be calculated in a quantum field theory through the replica
trick \cite{cc-04}
\be
S_A=\left.-\frac{\partial}{\partial n} {\rm Tr}\,\rho_A^n\right|_{n=1}\,.
\ee
This is particularly useful in CFT because, when $A$ consists of
disjoint intervals with $N$ boundary points with $B$ we have that 
${\rm Tr}\,\rho_A^n$ transforms under a general conformal transformation
as the $N$ point function of a primary field $\Phi_n$ with conformal
dimension \cite{cc-04} $\Delta_n= {c}/{12} (n-1/n)$, 
where $c$ denotes the central charge of the underlying CFT.
The fields $\Phi_n$ are usually called twist operators.
In particular this implies that the entanglement entropy of a slit
of length $\ell$ in an infinite system is given by \cite{ee,cc-04}
\be
{\rm Tr}\,\rho_A^n= c_n \left(\frac\ell{a}\right)^{-2\Delta_n}
\qquad\Rightarrow\qquad S_A=\frac{c}3 \log \frac{\ell}{a}- c'_1\,,
\ee
where $a$ is an UV cutoff (e.g. in a spin chain is the lattice spacing)  
and $c'_1$ a non universal constant.

Clearly within this mapping to correlation function we can calculate
the time-dependence of the entanglement entropy. For the case of homogenous
quenches this has been done in Ref. \cite{cc-05}, finding
\be
S_A(t) \simeq \cases{ \frac{\pi ct}{6\tz}       & $t<\ell/2$ \,,\cr
\frac{\pi c\,\ell}{12\tz} &$t>\ell/2$ \,. }
\ee
This result can be easily explained in terms of the causal scenario longly 
discussed there, and it is due to pairs of coherent quasi-particles emitted 
from any point and reaching one the subsystem $A$ and the other $B$.
In the case of $A$ and $B$ corresponding to two semi-infinite lines is 
a straightforward generalization of the previous result. In fact, we have only
to consider the one-point correlation function of a twist operator
$\langle\Phi_n\rangle$ that in the upper half-plane is 
$\langle\Phi_n(z,{\bar z})\rangle_{\mathtt U}= \tilde{c}_n [{\rm Im}
z]^{-\Delta_n}$ (where $\tilde{c}_n$ is the analogous of $c_n$ for the boundary
CFT introduced in \cite{cc-04}).
Mapping to the strip, analytically continuing, and taking the limit of large
time, directly gives that the entanglement entropy grows indefinitely with 
time as $S_A(t)=\pi ct/{12\tz}$.
Let us mention that this indefinite grow of entanglement is the main reason 
why numerical simulations (e.g. \`a la density matrix renormalization group)
do not perform efficiently for this kind of time evolution \cite{eff-dmrg}.

The natural question arising is how this result is
modified by the presence of the domain-wall at $r=0$.
Let us start with the case of two semi-infinite lines with $A=[r,\infty)$ and
$B=\mathbb{R}\backslash A$ its complement.
In this case the entanglement is given by the one-point function 
$\langle \Phi_n(r,t)\rangle $ of a twist operator at position $r$. 
From the general result reported above Eq. (\ref{1ptgen}), we know that
$S_A(t)$ increases linearly with time for $t<|r|$, as expected because the
time evolution is the same as in the absence of the domain-wall.
For $t>|r|$ we also find from Eq. (\ref{1ptgen}) a linear increase of the 
entropy, but its rate in principle could change if 
$f_{ab}^{\Phi_n}(0)$ is vanishing. 
However, the properties $\Delta_1=1$ and $\Tr\rho=1$ trivially leads to
${\cal A}_{ab}f_{ab}^{\Phi_1}(0)=1$. Analyticity in $n$ and $\cos \theta$ near 
zero makes then $f_{ab}^{\Phi_n}(\cos \theta \to 0 )$ non zero close to $n=1$ 
and this ensures that the domain-wall does not change the linear increasing 
rate of the entanglement entropy, but only changes it up to an additive 
constant.
The finite change, with respect to e.g. a homogeneous initial conditions of
type $a$, is given by 
${\cal S}_{ab}=-\partial_n {(\cal A}_{ab} (f_{ab}^{\Phi_n}(0)-f_{ab}^{\Phi_n}(1))|_{n=1}$,
independent of $r$ with $|r| < t$. 
Hence the entanglement of the two parts being initially separated 
by the domain-wall increases linearly with only {\it a universal additive 
constant} ${\cal S}_{ab}$ distinguishing from the homogeneous quench. 
This constant is computed below, in the particular cases relevant for SLE.

For the entanglement entropy of a finite slit $A=[r_1,r_2]$ with
$\ell=|r_1-r_2|$, general features are simple consequences of the scaling 
form in Eq. (\ref{2ptgen}).
As long as $t<|r_1|,|r_2|$, the time evolution remains the same as in the 
absence of the domain-wall (i.e. initial linear increase and then saturation).
Conversely for large times $t>|r_1|,|r_2|,\ell/2$ the entanglement entropy
saturates to a value proportional to $\ell$.
In the intermediate regimes, that depend on the nature of the domain-wall, 
from the general scaling again only finite universal shifts 
compared to the homogeneous quench can be possible.

\section{Connection with SLE}
\label{sec6}

Geometrical and statistical properties of a two-dimensional
critical system in the half-plane with an interface starting at the boundary
are nowadays widely studied by means of
the so called stochastic Loewner  evolution (SLE) that allows for a rigorous
mathematical derivation of some CFT results (see e.g. Refs. \cite{sle-rev}
as reviews). We do not enter here in the beautiful technical details of the approach, but
we will limit to state some results and ask whether they may have some
connection with the quench problem considered here. An SLE interface 
receives a CFT interpretation as being created by inserting a boundary changing operator
$\Phi_{12}$ at the boundary. In some cases, such as Ising, it corresponds to a domain wall
and an $ab$ boundary change. Hence for simplicity we stick here to the
Ising notations, keeping in mind that in this section we deal with boundary
condition changes relevant for SLE$_\kappa$ at generic value of the parameter $\kappa$. These may be quite
different from Ising, and are detailed in e.g. Refs. \cite{sle-rev}. 

One of the most celebrated result concerns the rigorous derivation
of Schramm's formula \cite{s-00}. Starting from the boundary the interface
explores the full upper half-plane
and ends at infinity. This curve (called the SLE trace) divides the 
half plane in two parts. Schramm's formula gives the probability that one point lies to the right
of this curve. In the continuum limit, this probability only depends on the
angle $\theta$ that the point makes with the boundary (that is the same
$\theta$ we used in all the paper). Schramm's formula gives this
probability as
\be
p(\theta)= H(\cot \theta) = \frac12
 + \frac{\Gamma(4/\kappa) \cot\theta }{\sqrt{\pi}\,\Gamma{(4/\kappa-1/2)}}
{}_2{\rm F}_1\left(\frac{1}{2},\frac{4}{\kappa},\frac{3}{2};
-\cot^2\theta\right)\,,
\ee
satisfying the ``boundary conditions'' $p(\theta = \pi/2)=1/2$ (reflection
symmetry) and $p(\theta =\pi)=0$ (point at the negative real axis).
The parameter $\kappa$ is the diffusion constant of the underlying stochastic
process and turns out to define different universality classes (e.g. the Ising
model corresponds to $\kappa=3$) and is related to the central charge.

Since SLE focuses on geometrical properties of interfaces, one could try to 
push the  continuation to real time $t$, and define a space-time interface
geometrically using the path integral  formulation. This can be done in
principle by appropriate continuum limit (in the time direction) of lattice
models such that the interface remains well defined. Consider the Ising case
as a typical example. 
Specifying an interface means specifying the local spins $+$
(to the left) and $-$ (to the right) along a simple curve in the $r,t$ plane
(since it wanders this entails creation and annihilation of domain-walls as
$t$ increases). One must then superpose the complex amplitudes associated with
all configurations satisfying that dynamical constraint. These are
continuations of the real positive probabilities used to construct
SLE. Schramm's dimension zero operator additionally constraints the interface
to pass right of a space time point (i.e. to be connected to one of the two
boundary conditions $a$ or $b$ without crossing the interface). Clearly this
qualitative picture needs to be made more concrete. 

\begin{figure}
\centering
\includegraphics[scale=1.1]{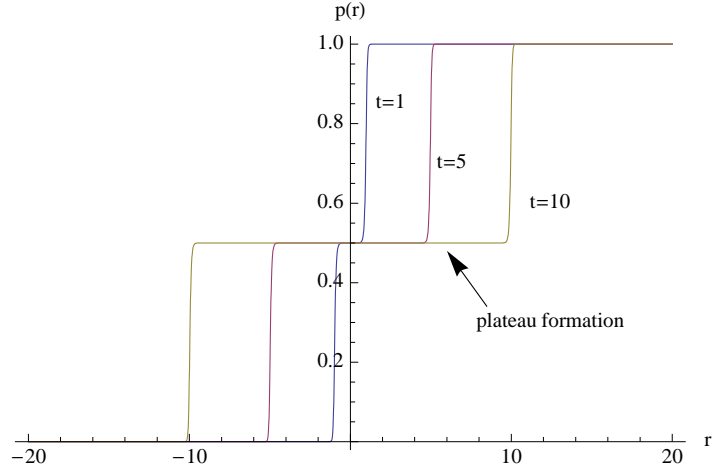}
\caption{Probability $p(r,t)$ as given in
  Eq. (\ref{eqn:proba}) for different times $t$. Here, the parameters are
  $\tau_0 =0.1$ and $\kappa=3$ (Ising model).
  At large times, a plateau around $r=0$ at $p=1/2$ emerges. }
\label{fig:probs}
\end{figure}

Can $p(\theta)$ be used to determine some properties of the quench
problem studied here? In order to address this question,
let us map $p(\theta)$ to the strip according to
Eq. (\ref{map}),
and then perform the analytical continuation $\tau \to \tau_0+i t$.
A simple calculation gives
\be
p(r,t) =
H\left(\cot\theta= \frac{\sinh(\pi r/2\tz)}{\cosh(\pi t /(2 \tz) )} \right)\,,
\label{eqn:proba}
\ee
that is a positive number smaller than $1$. Hence it could in principle still be interpreted
as a probability. 
In the limit $t,|r|\gg\tz$,
using the properties of the hypergeometric function, we find
\begin{equation}
  p(r,t) \stackrel{|r|,t\gg\tz}{=}
  \left\{
  \begin{array}{rc}
  0& r < -t,\\
  1/2& -t < r < t,\\
  1& r > t,
  \end{array}
  \right.
\end{equation}
independently from $\kappa$. This result is suggestive and 
could be thought, in a sense which remains to be made precise,
as a probability of being in the state evolved from $+$, that is clearly zero
for $r<-t$ and $1$ for $r>t$. A plateau develops (see figure \ref{fig:probs}) 
for $r < |t|$ and by mirror symmetry it can here only take the value $1/2$. It
would be interesting to understand how this plateau forms from the path
integral formulation, presumably from interference effect. From the point of
view of a real time relativistic theory it is however rather natural.

Are there other probabilities to which this could be compared? A priori the
quantum state evolution is fully deterministic and apparently there is no 
probability into the game.
A possible way to introduce a probability proceeds as follows. Consider, in a
lattice model, the reduced density matrix $\rho_r(t)$ of the degrees of
freedom localized in $r$ (i.e. for a spin chain the spin in $r$).
Clearly for $r>t$, $\rho_r(t)$ equals the analogous one obtained from the
evolution of a homogenous $+$ boundary condition $\rho^+_r(t)$.
Analogously for $r<-t$, $\rho_r(t)$ is $\rho^-_r(t)$. An interesting
possibility is that the reduced density matrix in the center region, 
$-r<t<r$, is simply a combination of the above two, in the present
case by symmetry it can only be $\rho_r(t)=(\rho^-_r(t)+\rho^+_r(t))/2$.
An interesting question then is whether the $1/2$ found from Schramm's formula is the
same factor appearing in front of the $\rho^+_r(t)$ reduced density matrix. It is
easy to check in the Ising case, from our result for the averaged spin, 
that this could only hold strictly for infinite time (the leading
decaying term in $e^{-\pi t/2\tz}$ {\it does not} agree). 
Unfortunately, in that limit,
the factor $1/2$ being imposed by symmetry does not provide a test.
Tests would require more complicated situations, e.g. translating the
results of Refs. \cite{gc-05,drc-05,bbk-05} in terms of quantum evolution.
For instance, in Ref. \cite{gc-05} the probability 
$P_{middle}(\cot \theta)$ was obtained,
that a point $z$ in the upper half plane
lies in between (left of one, right of the other) two interfaces starting at nearby points
on the real axis, {\it conditioned} to the fact they do not cross (i.e. going to infinity).
This may be relevant to determine whether, in presence of {\it two} 
domain-walls in the initial state, e.g. an interval $[0,r_0]$ of $-$ spins in 
an otherwise homogeneous $+$ state,
the point $r\gg r_0$ at time $t$ still feels the presence of the $-$ in the
initial condition (however, this provides only part of the answer, since one
should add the event where the domain-wall annihilates at late times, but
this  also can be obtained, in principle, within SLE). 
The large time continuation yields
$P_{middle}(\cot \theta) \to D_\kappa$, the constant defined in 
Ref. \cite{gc-05}, for all $t>|r|$, which has a non trivial value,
e.g. $D_{\kappa=3}=0.921866$ for Ising.

To close on a less speculative note, let us give the explicit form for the 
one-point function of any primary field of scaling dimension $\Delta$ in
presence of the boundary conditions which are relevant for SLE. It is known
that the boundary changing operator which creates an SLE$_\kappa$ interface is
$\Phi_{12}$, with conformal dimension $\Delta_{12}=2h_{12}= (6-\kappa)/\kappa$.
The result of Ref. \cite{bx-90} is then Eq. (\ref{1ptgen}) with
\begin{equation}
f_{ab}(\cos\theta)= C_1\, {_2}F_1(\alpha,\beta,1/2,\cos^2\theta)+C_2 \cos\theta \, {_2}F_1(\alpha+1/2,\beta+1/2,3/2,\cos^2\theta)\,,
\end{equation}
where for convenience we have absorbed the amplitude ${\cal A}_{ab}$ into 
$f_{ab}$, with parameters
\begin{equation}
\fl\alpha,\beta=
\frac{1}{6}\left(1-2\Delta_{12}\pm\left[(1-2\Delta_{12})^2+
12 \Delta (1+\Delta_{12})\right]^{1/2}\right)=
(\kappa -4  \pm \sqrt{(\kappa-4)^2 + 8 \kappa \Delta )})/(2 \kappa)\,. 
\label{alphabeta} 
 \end{equation}
We determine the constants $C_1$ and $C_2$ from the boundary conditions
$f_{ab}(\pm 1)$ at $\theta=0,\,\pi$ (respectively to the right/left of the
domain-wall), that yields
\begin{equation}
C_1 =\frac{1}{2} \frac{f_{ab}(1)+f_{ab}(-1)}{ {_2}F_1(\alpha,\beta,1/2,1)}\,, 
\qquad
C_2= \frac{1}{2} \frac{f_{ab}(1)-f_{ab}(-1)}{ {_2}F_1(\alpha+1/2,\beta+1/2,3/2,1)}\,. \label{c1c2} 
\end{equation}
Using the identity 
${_2}F_1(a,b,c,1) = \Gamma(c)\Gamma(c-a-b)/(\Gamma(c-a)\Gamma(c-b))$,
we find $c-a-b= (8-\kappa)/2\kappa$. 
Hence both denominators are finite for $0<\kappa<8$ 
(the denominator associated to $C_1$ equals $1$ for all $\kappa$ for
$\Delta=0$ and this is consistent with the result for the twist operators with
$\Delta_1=0$ in the entanglement entropy section). 
However we find that $C_1$ has a pole for  $\Delta=\Delta_n^{(1)} =
\frac{\kappa}{2}\left(n+\frac{1}{2}\right)\left(n+\frac{4}{\kappa}-\frac{1}{2}\right)$
and $C_2$ for $\Delta_n^{(2)}
=\frac{\kappa}{2}\left(n+1\right)\left(n+\frac{4}{\kappa}\right)$ both for
non-negative integer $n$ (note that the latter vanishes for $\Delta=2$ for all
$\kappa$). 
Hence for such operators the symmetry must be imposed for the result to be
meaningful. Obviously one recovers the Ising results Eqs. (\ref{sigIs}) and
(\ref{eIs}) for $\kappa=3$ inserting respectively $\Delta=1/8$ (spin) and
$\Delta=1$ (energy).

Using the continuation formula (\ref{cosf}) one finds formally the large time
asymptotics for $t > |r|$ is given by Eq. (\ref{onepointgen}) with 
\begin{equation}
{\cal A}_{ab} f_{ab}(0) \to  C_1 + C_2 e^{-\pi(t-|r|)/2\tz} + \dots\,,
\end{equation}
where the second decaying term becomes the leading one when $C_1=0$, else
it is a rapidly decaying subleading correction. 

Note that the Schramm formula is recovered, for some choice of boundary 
condition, considering an operator with scaling scaling dimension $\Delta=0$. 
One gets then $\beta=0$ and $\alpha= 1-4/\kappa$ and
$$
f_{ab}(\cos\theta)= 
C_1 + C_2\cos \theta \, {_2}F_1(1/2,3/2-4/\kappa,3/2,\cos^2\theta)
= C_1 + C_2 \cot \theta \,{_2}F_1(1/2,4/\kappa,3/2,-\cot^2\theta)\,,
$$
using properties of hypergeometric functions. It remains to be seen whether a 
dimension zero primary operator, relevant for the quench, can be identified. 
In that case it should be related to Schramm formula (or be trivial). 

We can now compute, in the case of an SLE interface, the universal additive 
contribution ${\cal S}_{ab}$ to the entanglement
entropy as discussed in Section \ref{sec5}. Inserting $\Delta=c /12 (n-1/n)$ which yields the appropriate values for $\alpha_n$ and $\beta_n$, and using the symmetry $f_{ab}^{\Phi_n}(-1)= f_{ab}^{\Phi_n}(1)$, we find
\begin{equation}
{\cal S}_{ab} = - \partial_n [{_2}F_1(\alpha_n,\beta_n,1/2,1)]^{-1} |_{n=1} = c \frac{\psi(\frac{4}{\kappa} - \frac{1}{2}) +\gamma_E + \ln 4}{3 (\kappa-4)}\,, 
\label{Sab}
\end{equation}
where $\psi(x)$ stands for the Polygamma function and, in terms of $\kappa$, 
the central charge is $c=(6-\kappa)(3 \kappa - 8)/(2 \kappa)$. 
Notice that for $\kappa<4$ (and in particular for the Ising model) this
universal shift is negative.

\subsection{Two domain-walls}

\begin{figure}[htpb]
  \centering
  \begin{pspicture}(0,0.5)(7,5)
    \psframe[fillstyle=solid,fillcolor=lightgray,linewidth=0](1,1.3)(7,1.5)
    \psframe[fillstyle=vlines,fillcolor=lightgray,linewidth=0](1,1.3)(3,1.5)
    \psframe[fillstyle=hlines,fillcolor=lightgray,linewidth=0](3,1.3)(5,1.5)
    \psframe[fillstyle=vlines,fillcolor=lightgray,linewidth=0](5,1.3)(7,1.5)
    \psframe[fillstyle=solid,fillcolor=lightgray,linewidth=0](1,4)(7,4.2)
    \psframe[fillstyle=vlines,fillcolor=lightgray,linewidth=0](1,4)(3,4.2)
    \psframe[fillstyle=hlines,fillcolor=lightgray,linewidth=0](3,4)(5,4.2)
    \psframe[fillstyle=vlines,fillcolor=lightgray,linewidth=0](5,4)(7,4.2)
    \psline[linewidth=1.5pt](1,1.5)(7,1.5)
    \psline[linewidth=1.5pt](1,4)(7,4)
    \psline[linestyle=dotted](0.8,2.75)(7,2.75)
    \psline{->}(0.8,1.3)(0.8,4.5)
    \psline{->}(1,0.8)(7,0.8)
    \psline(4,0.7)(4,0.9)\psline(3,0.7)(3,0.9)\psline(5,0.7)(5,0.9)
    \rput(3,0.5){$-\ell/2$}\rput(5,0.5){$+\ell/2$}
    \rput[c](0.5,1.5){$0$}\rput[c](0.5,2.75){$i\tau_0$}\rput[c](0.4,4){$2i\tau_0$}
    \rput[c](0.5,4.75){$i\tau$}
    \psline(0.75,1.5)(0.85,1.5)\psline(0.75,2.75)(0.85,2.75)\psline(0.75,4)(0.85,4)
    \rput(4,0.5){$0$}\rput(6.9,1.1){$r$}
    \rput(2,1.1){$a$} \rput(6,1.1){$a$}\rput(4,1.1){$b$}
    \rput(2,4.4){$a$} \rput(6,4.4){$a$} \rput(4,4.4){$b$}
    \psdot(5.3,2.75)\rput[c](5.6,3.2){$w=r+i\tau$}
    \rput(1.5,1.8){\Large ${\mathtt S}$}
  \end{pspicture}\hspace{1.cm}
  \caption{Strip geometry with domain-walls at $\pm \ell/2$ and $\pm \ell/2 + 2i\tau_0$.}
  \label{fig:stripblock}
\end{figure}
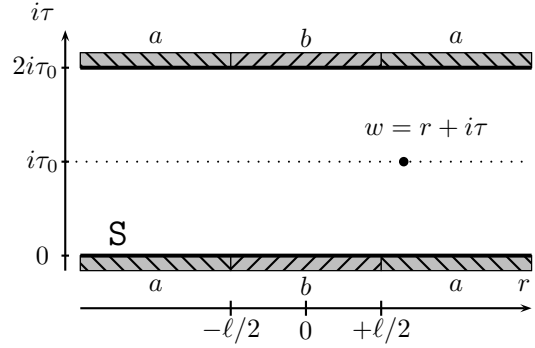

Here we consider the case of two domain-walls in the initial state. It is more general, but in the context of
SLE, it corresponds to a two interface problem and can be
studied, for appropriate boundary conditions,
by insertion of two $\Phi_{12}$ operators on each side in the strip. 
The corresponding strip geometry is depicted in Fig. \ref{fig:stripblock}.
However, the complete correlation functions mapped on the upper half-plane
need the insertion of four changing operators, that are in principle
calculable but very cumbersome. To make progress, we first consider the limit
of a small $b$ domain in a
line $a$ when results can be obtained from the 
fusion of two $\Phi_{12}$ operators, which produces $\Phi_{13}$ operators.
At the end, we finally argue that for asymptotic large time the result is valid for any finite fixed
size domain $b$, not necessarily small. 
According to physical intuition this means that no matter how big
is the $b$ domain, it is always small compared to the infinite remaining $a$
part. 

Let us consider the operator $\mathcal{O}=\prod_i\Phi_i(w_i,\overline{w}_i)$
being a product of primary operators. 
When $|x_1-x_2|=|x_3-x_4|=\ell$ is small, taking into account the insertion of
boundary condition changing operators, and the (generic) fusion algebra
$\Phi_{12}\Phi_{12}\sim 1+\Phi_{13}$, we obtain 
\begin{eqnarray}
\fl\langle \mathcal{O} \rangle^{\mathtt S}_{aba}&=& \ell^{4\Delta_{12}}
\left\langle \mathcal{O} \prod_{k=1}^4\Phi_{12}(x_k)\right\rangle_{\mathtt S}=
\nonumber\\
\fl&=& 
\langle \mathcal{O} \rangle^{\mathtt S}_a
+ \ell^{\Delta_{13}} (\langle \mathcal{O} \, \Phi_{13}(0)\rangle^{\mathtt S}_a+
\langle \mathcal{O} \, \Phi_{13}(2i\tau_0)\rangle^{\mathtt S}_a )  + \ell^{2 \Delta_{13}}
\langle\mathcal{O}\,\Phi_{13}(0)\Phi_{13}(2i\tau_0)\rangle^{\mathtt S}_a 
+ \dots\,, \label{expansion13} 
\end{eqnarray}
where the dots denote the contribution of descendant operators, which are
easily seen to include analytic $O(\ell^2)$ corrections as compared to the
leading term which corresponds to homogeneous boundaries. 
The dimension of the $\Phi_{13}$ operator is, in the SLE context:
\be
\Delta_{13} = \frac{8-\kappa}{\kappa}\,. 
\ee
In terms of SLE of each term in Eq. (\ref{expansion13}) can be interpreted as
follows. The second term represents the contribution of two interfaces
starting both very close to zero (respectively to $2 i \tau_0$) on the same
side of the strip and which do not immediately annihilate near their origin
(but rather annihilate in the bulk). The last term corresponds to the
possibility where the two interfaces cross the whole strip. It has thus some
general relation to the calculation of Ref. \cite{gc-05}, relation which can be made
precise for a dimension zero operator. We recall here that we use Ising notations for convenience only
but that we consider generic $\kappa$ and the appropriate boundary changes \cite{sle-rev}). 

Let us now specialize to a single primary field
$\mathcal{O}=\Phi(w)$ of conformal dimension $\Delta$, being
inserted at $w=r + i\tau$. 
As explained above, at the leading order in $\ell$, these can be obtained 
simply from $\langle \Phi \rangle^{\mathtt S}_a$, 
$\langle \Phi \Phi_{13}(0)\rangle^{\mathtt S}_a$, and
$\langle \Phi \Phi_{13}(0)\Phi_{13}(2i\tau_0)\rangle^{\mathtt S}_a$.
To calculate this one should map the result in the upper half-plane, that are
obtained again by fusing two $\Phi_{12}$. 
The calculations are straightforward. For example, applying a Moebius map
to Eq. (\ref{onepointgen}) one obtains the one point average in presence of a single finite domain, $\langle \Phi(z)\rangle^{\mathtt U}_{aba}= \ell^{2 \Delta_{12}} \langle \Phi(z) \Phi_{12}(0) \Phi_{12}(l) \rangle_{\mathtt{U}}
= y^{-\Delta} f^{\Phi}_{ab}\left(\cos \psi\right)$ with
$\cos \psi = {\rm Re}(\sqrt{z (l-\bar z)/\bar z (l- z)})$ 
(the constant ${\cal A}_{ab}$ is absorbed in $f^{\Phi}_{ab}$ again).
This gives the lowest order terms in Eq. (\ref{expansion13}), with the
correct amplitudes, from the expansion of 
$f^{\Phi}_{ab}(\eta=1-\zeta)/f^{\Phi}_{ab}(1)=1+C_a \zeta^{\Delta_{13}/2}+ B_a \zeta +\dots$ for small
$\zeta=\frac{1}{2} (\ell/y)^2 \sin^4 \theta + O(\ell^4)$. 
It yields in particular $ \langle \Phi(z)\Phi_{13}(0)\rangle_{\mathtt{U}}=
C_a(\sin\theta)^{2 \Delta_{13}}y^{-\Delta -\Delta_{13}}$. 
Putting everything together we finally obtain
\be
\fl\langle \Phi \rangle^{\mathtt U}_{ababa}\sim
\langle \Phi \rangle^{\mathtt U}_a \left(1+C_a 
\left(\frac{\ell}{y}\right)^{\Delta_{13}} (\sin\theta)^{2 \Delta_{13}}+ B_a \left(\frac{\ell}{y}\right)^{2} (\sin \theta)^4 + 
\left(\frac{\ell}{y}\right)^{2\Delta_{13}} g^\Phi(\cos\theta)+ ..
\right)\,,
\label{final} 
\ee
with $B_a = \alpha \beta/(1/2+\alpha + \beta)=4 \Delta/(8 - 3 \kappa)$ and 
$C_a= \hf{1/2-\alpha}{1/2-\beta}{1/2}{1} (C_1/f_{ab}(1)) 
+\hf{1-\alpha}{1-\beta}{3/2}{1} (C_2/f_{ab}(1)) $, where $C_1$ and $C_2$ are
given in (\ref{c1c2}), $\alpha,\beta$ in (\ref{alphabeta}), and the relation
given below (\ref{c1c2}) to ratio of Gamma function should be used to define
the relevant limits. The general expression for the function
$g^\Phi(\cos\theta)$ is given in Ref. \cite{bx-90} in terms of hypergeometric
functions, using the fact that $\Phi_{13}$ being degenerate at level three,
$\langle \Phi(z) \Phi_{13}(0)\Phi_{13}(\infty)\rangle^{\mathtt U}_a$ obeys a
simple differential equation. Note that the descendants will give non trivial
corrections to (\ref{final}), but these are always subdominant to the leading
correction and, quite often subdominant even to subleading ones. Hence in the
general expansion in small $(\ell/y)$ at fixed $\cos \theta$, the leading
correction comes from the $(\ell/y)^{\Delta_{13}}$ term only for
$\Delta_{13}<2$, i.e. for SLE type boundary conditions, $\kappa > 8/3$. 
For $\kappa < 8/3$ the $B_a$ term dominates. 
Note however that even for $\kappa > 8/3$ the $B_a$ term dominates if the 
amplitude vanishes $C_a=0$. 
This is the case for the Ising model $\kappa=3$ if $\Phi$ is
either the bulk spin or energy operator, since $f_{ab}^\Phi(\eta)$ is
analytic in these cases as discussed in Section \ref{sec4}. 
The last term is often subleading, although
in principle there may be cases, for $\kappa > 4$, where $C_a$ vanishes and
the leading correction is given by the last term. Another example consists in the
$O(n)$ loop model when each interface carries a different $O(n)$ index hence they cannot annihilate. 
The general case depend on
the operator $\Phi$, on the model, and on $\kappa$, and can be read from the above results.

Mapping to the strip and continuing to the real time quantum dynamics, we know 
that for $t>|r|$ the half plane variable
$(\ell/y) \sin^2 \theta$ in (\ref{final}) maps to $\ell e^{-\pi (t-r)/2\tz}$, a variable which is small {\it both} in the
small $\ell$ and the large time limit. Hence we find for large times $t>|r|$ 
\be
\langle \Phi(r,t) \rangle_{aba}=
\langle \Phi(r,t) \rangle_a 
(1+ C_a(\ell e^{-\pi (t-r)/2\tz})^{\Delta_{13}} + B_a \ell^2 e^{-\pi (t-r)/\tz} + \dots)\,. \label{resfin} 
\ee
where one of the two correction terms dominates depending on whether $\Delta_{13}$ is larger or smaller 
than $2$, and of whether $C_a$ vanishes or not, as discussed above. 
The scaling variable governing the correction is then $\ell e^{-\pi t/2\tz}$,
so no matter how large $\ell$ is, provided it is finite, for long times the
leading correction is always of the above form. Physically this corresponds to
the property that asymptotically 
the $b$ domain is washed out and the time evolution is the same as from an
homogeneous $a$ initial condition with only exponentially small
corrections. Although we have not attempted any calculation it is physically
clear that any initial state with a local domain structure immersed in an
infinite $a$ initial state will behave in a similar fashion, possibly
involving a spectrum of higher $\Delta_{1,2 N+1}$ exponents, 
corresponding to $2 N$ SLE curves. Applying to $\Delta \to \Delta_n$ and
following the arguments of Section \ref{sec5}, one also easily sees that the
difference in entanglement entropy as compared to the homogeneous case decays
exponentially in time, with the same decay rates as in (\ref{resfin}) and
prefactors given by $- \partial_n (C_a,B_a)|_{n=1}$. Explicit calculation
shows that the leading amplitude does not vanish. Similarly, in presence of an
{\it odd} number of domain walls in a finite region in the initial state, the
difference in entanglement entropy as compared to the homogeneous state, is
expected to converge at large time, with similar exponentials, to the value of
${\cal S}_{ab}$ given by (\ref{Sab}).

\section{Conclusions}
\label{sec7}

\begin{figure}[htpb]
\centering
\includegraphics[width=0.7\textwidth]{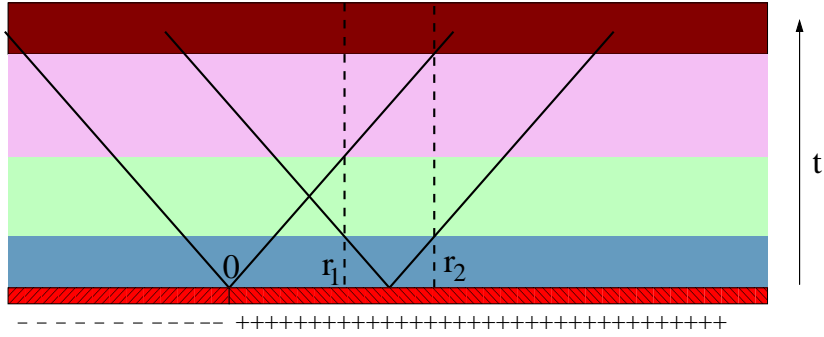}
\caption{Causal interpretation of the results for all the two-point functions
$\langle O(r_1,t) O(r_2,t)\rangle$.
In the case shown with $(r_2-r_1)/2<r_1<r_2$ and $r_i>0$, there are four
different time regimes, shown with different colors in the figure.
1) For $2t<r_2-r_1$, the two points evolve independently (blue region).
2) For $(r_2-r_1)/2<t<r_1$, the two-point function is quasi stationary because
there is not yet an influence of the domain-wall (green region).
3) For $r_1<t<r_2$, the domain-wall enters in the dynamics (violet).
4) For $t>r_2$, stationary behavior with quasi-thermal correlations (red).
}
\label{fig:cones}
\end{figure}

We presented a detailed study of the time evolution of a 1D quantum system
evolving according to a gapless Hamiltonian from an initial state with one
domain-wall. All our findings can be easily interpreted in terms of
quasi-particles excitations emitted at time $t=0$, and then propagating trough
the system at a finite speed given by the sound velocity $v_s=1$.
According to this scenario any correlation function has different time
regimes.
For example, in figure \ref{fig:cones} the time evolution  of the two-point
correlation functions has different regimes resulting from the competition
of the quasi-particles emitted from the middle point and those emitted from
the domain-wall. The same interpretation is also valid for the entanglement
entropy. 

The importance of this simple physical picture is that it easily allows to 
understand  limit and generality of the CFT results. In fact, the behavior 
outside the causal cones is
only consequence of the finite velocity of sound and so it must be generally
true also for  gapped systems, as pointed out in details in 
Refs. \cite{cc-06,cc-07}. Oppositely, lattice models, even in the
thermodynamic limit, have a full spectrum for the velocity of excitations, due
to non-linear dispersion relations. These slow quasi-particle give corrections
inside the causal cone to the results just derived (see Ref. \cite{cc-07} for a
careful discussion of this  issue).

As in the case of an homogeneous quench, we showed that for large times all 
the correlation functions are the same as in a thermal state with effective
temperature $\beta_{\rm eff}=4\tz$. This means that the presence of the
domain-wall does not affect the asymptotic state. In fact, this feature does 
not come unexpected since the energy of this state is only slightly larger 
than the translational invariant one. Following this interpretation, 
we conclude that a state with an infinite number of domain-walls should change 
the effective temperature.

We finally discussed the possibility that the geometric picture obtained
from SLE may be continued from imaginary to real time. Whether this 
may lead to new information for quantum quenches is not yet clear at this 
stage, but has been critically discussed in the present paper. We have also
obtained, for the type of domain wall considered in SLE, the universal change
in entanglement entropy as compared to an homogeneous initial state. Finally
we considered the case of few domain walls in a finite region, and obtained the
leading time decay of one point functions.

\section*{Acknowledgments}
PC thanks John Cardy, Dragi Karevski, and Erik Tonni for useful discussions. This work has been done in part
when PC was a guest of the Institute for Theoretical Physics of the Universiteit van Amsterdam (a stay supported
by the ESF Exchange Grant 1311 of the INSTANS activity) and in part as guest at Ecole Normale Superieure whose
hospitality is kindly acknowledged. CH benifits from financial support from the French \textit{Minist\`ere de
l'Education et de la Recherche} and PLD from ANR program 05-BLAN-0099-01.

\appendix

\section{Useful formulas for the analytic continuation}
\label{appA}

The strip geometry is obtained by the conformal mapping of the upper half-plane $z=x+iy$ to the strip $w=r+i
\tau$ given by Eq. (\ref{map}). It is useful to write it explicitly in cartesian and polar coordinate. We have
\bea z&=& \frac{e^{\pi w /(2 \tz)}-1 }{e^{\pi w /(2 \tz)} + 1} =
\tanh\left(\frac{\pi w}{4\tz}\right)\,, \nonumber \\
x&=& \mathrm{Re}\,z =
\frac{\sinh(\pi r/ 2\tz)}{\cos(\pi \tau/2 \tz) +\cosh(\pi r/2 \tz)}\,, \nonumber\\
y&=& \mathrm{Im}\,z =
\frac{\sin(\pi \tau/2\tz)}{\cos(\pi \tau/2 \tz) +\cosh(\pi r/2 \tz)}\,, \nonumber\\
|z|^2&=& \frac{\cosh(\pi r/\tz) -\cos(\pi\tau/\tz)}{2[\cos(\pi\tau/2 \tz) +\cosh(\pi r/2 \tz)]^2}\,. \eea Upon
continuation $\t\to \tz+it$ one has $\sin(\pi \tau/2 \tz) \to \cosh(\pi t/(2 \tau_0))$ and $\cos(\pi \tau/ \tz)
\to - \cosh(\pi t/\tau_0)$, but $\cos(\pi \tau/(2 \tz))$ continues to an imaginary value. Hence the denominator
must disappear in any physical quantity, and it does, being compensated by the
Jacobian
\bea |z'(w)|=
\frac{\pi}{4 \tau_0} |1-z(w)^2| = \frac{\pi}{2 \tau_0 (\cos(\pi\tau/2 \tz) +\cosh(\pi r/2 \tz))} = y
\frac{\pi/(2 \tau_0)}{\sin(\pi \tau/2\tz)}\,. 
\eea

The polar angle $\theta$ is mapped to the strip according to
\begin{equation*}\fl
\cos\theta=\frac{x}{|z|}=
\frac{\sqrt2\sinh(\pi r/2\tz)}{
[\cosh(\pi r/\tz)-\cos(\pi\tau/\tz)]^{1/2}}\,,
\qquad
\sin\theta=\frac{y}{|z|}=
\frac{\sqrt2\sin(\pi \t/2\tz)}{
[\cosh(\pi r/\tz)-\cos(\pi\tau/\tz)]^{1/2}}\,,
\end{equation*}
After analytic continuation we have \bea \cos\theta&\to& \frac{\sqrt2\sinh( \pi r/2\tz)}{[\cosh(\pi r / \tz)
+\cosh(\pi t/ \tz)]^{1/2}} \simeq {\rm sign}\, r \times \cases{ 1                    & $t<|r|$,\cr
        e^{-\pi(t-|r|)/2\tz} & $t>|r|$,
}\nonumber\\
\sin\theta&\to& \frac{\sqrt2\cosh( \pi t/2\tz)}{[\cosh(\pi r / \tz) +\cosh(\pi t/ \tz)]^{1/2}} \simeq \cases{
e^{\pi(t-|r|)/2\tz} & $t<|r|$,\cr 1 & $t>|r|$.
}\\
\tan\theta&\to&
\frac{\cosh \pi t/2\tz}{\sinh \pi r/2\tz}\simeq  {\rm sign}\, r\;
e^{\pi(t-|r|)/2\tz}\,,
\eea
where here and in the following $\simeq$ (and in the main text) stands
for $t,r\gg \tz$.

We will also need the following functions of the angles that are trivially
obtained from the previous formulae
\bea
\sin 2\theta&=&2\sin\theta\cos\theta\simeq 2\,
{\rm sign}\, r\; e^{-\pi|t-|r||/2\tz}\,,\\
 \cos\frac{\theta}{2}&=&\left(\frac{1+\cos\theta}2\right)^{1/2}\simeq
\cases{
\Theta(r)&$t<|r|$,\cr
2^{-1/2} &$t>|r|$,
}
\\
\tan\frac{\theta}{2}&=&\sqrt{\frac{1-\cos\theta}{1+\cos\theta}}\quad\simeq
  \cases{
    1 & $t>|r|$,\cr
    \displaystyle\frac{\Theta(-r)}{\Theta(r)} & $t<|r|$.
  }\\
4\cos^2\theta-3&=&\left(\frac{8 \sinh^2\pi r/2\t_0}{\cosh(\pi r/\tz) -\cos(\pi\t/\tz)}-3\right)\simeq
-1+2\,{\rm sign}\, (|r|-t)
\eea
Here, $\Theta(r)$ denotes the Heaviside function which is $1$ for $r>0$, and
$0$ otherwise. Note the divergence for $r<-t$ in $\tan\theta/2$. However,
this divergence cancels with the one-point function with free boundary
condition and we do not need the first order correction in $\tz$ that makes it
finite.

For the two-point functions, we need to study the combination of different
variables. In particular we need
\begin{equation}\fl
 u^4 = \left(1+\frac{4y_1y_2}{\rho^2}\right)=
\frac{(x_1-x_2)^2+(y_1+y_2)^2}{(x_1-x_2)^2+(y_1-y_2)^2}=
\frac{\cosh(\pi(r_1-r_2)/2\tau_0)- \cos(\pi (\tau_1+\tau_2)/2\tau_0)}{
\cosh(\pi(r_1-r_2)/2\tau_0)- \cos(\pi (\tau_1-\tau_2)/2\tau_0)}\,,
\end{equation}
that is easily analytical continued to $\tau_{1/2}\to \tau_0+it$ as 
\be
\fl u^4 = 1 + \frac{\cosh^2(\pi t/2 \tau_0)}{\cosh^2(\pi (r_1-r_2)/4 \tau_0)} = \cases{ 1 +  e^{ \pi
(2t-|r_1-r_2|)/2\tau_0}  \approx 1 & $|r_1-r_2|
>2t \gg\tau_0$,\cr e^{ \pi (2t-|r_1-r_2|)/2\tau_0}\gg 1 & $2t> |r_1-r_2| \gg\tau_0$. } \label{eqn:u4continued}
\ee
In particular all the terms like $(u^{2m}-1)$ vanish for $2t<|r_1-r_2|$
and the ratios $(u^{2m}-c_1)/(u^{2m}+c_2)$ tend to $1$
for $2t>|r_1-r_2|$, for all constants $c_i$.

For combinations of polar angles, we have 
\begin{equation*}
\fl \sin(\theta_1\pm \theta_2)=
\sin\theta_1\cos\theta_2\pm \sin\theta_2\cos\theta_1\simeq
\cases{
s_1 \,e^{-\pi(t-|r_1|)/2\tz} \pm s_2\, e^{-\pi(t-|r_2|)/2\tau_0} &
$|r_1|,|r_2|<t$,\cr
s_2 \,\pm s_1 \,e^{-\pi(|r_2|-|r_1|)/2\tau_0}& $|r_1|<t<|r_2|$,\cr
s_1 \,e^{\pi(t-|r_1|)/2\tau_0} \pm s_2
    \,e^{\pi(t-|r_2|)/2\tau_0}& $|r_1|,|r_2|>t$,
}
\end{equation*}
\begin{equation*}
\fl \cos(\theta_1\pm \theta_2)=
\cos\theta_1\cos\theta_2\mp \sin\theta_2\sin\theta_1\simeq
\cases{
s_1 s_2 \,e^{-\pi(t-|r_1|)/2\tz} e^{-\pi(t-|r_2|)/2\tz} \mp 1&
$|r_1|,|r_2|<t$,\cr
s_1 s_2 e^{-\pi(t-|r_1|)/2\tz}\mp e^{\pi(t-|r_2|)/2\tz}  & $|r_1|<t<|r_2|$\cr
s_1 s_2 \mp e^{\pi(t-|r_1|)/2\tz}  e^{\pi(t-|r_2|)/2\tz}& $|r_1|,|r_2|>t$,
}
\end{equation*}
where again $s_i={\rm sign}\, s_i$.

\section{The two-point function of different operators in the quench from a
  homogeneous initial state}
\label{appB}

In Refs. \cite{cc-06,cc-07} it has been shown that when the initial state is a
translation invariant state, the long-time correlation functions are the
same as a pseudo-thermal state with $\beta_{\rm eff}=4\tau_0$.
Then it follows that the correlation of operators with different scaling
dimensions must vanish, but we still do not know how they fall off to zero.
In this appendix we fill this small gap.

We first consider the energy-spin correlation in the Ising model starting from
a homogeneous initial condition (we only consider fixed, since free boundary
conditions gives $\sigma$ correlation identically zero).
In the upper half-plane ${\mathtt U}$ the correlation function is
\cite{m-87}
\be
\langle\sigma(1)\e(2)\rangle_+^{\mathtt U}=
\langle\sigma(1)\rangle_+^{\mathtt U}\langle\e(2)\rangle_+^{\mathtt U}
\frac{u^4+1}{2u^2}\,,
\ee
where $\langle\sigma(1)\rangle_+^{\mathtt U}\propto y_1^{-1/8}$, and
$\langle\e(2)\rangle_+^{\mathtt U}\propto y_2^{-1}$, and $u$ is defined as in (\ref{eqn:defu})
(see \ref{appA} for an explanation how to manipulate it).
For $2t<|r_1-r_2|$, we have $u\sim 1$ and the correlation function is
the product of the one-point functions, in agreement with the causal
interpretation.
Oppositely when $2t>|r_1-r_2|$, because $u^4\sim e^{\pi(2t-|r_1-r_2|)/2\tz}$,
we have
\be
\langle\sigma(1)\e(2)\rangle_+\propto
e^{-\pi t/16\tz}e^{-\pi t/2\tz} e^{\pi (2t-|r_1-r_2|)/4\tz}=
e^{-\pi t/16\tz}e^{-\pi |r_1-r_2|/4\tz}\,,
\ee
i.e. the two-point function is going to zero with an exponential rate.

Another simple example is the Gaussian model defined by the continuum
Hamiltonian $H\propto \int_{y>0} (\nabla \phi(z))^2 dx dy$.
The field $\phi$ is not primary, but its exponential $e^{i\alpha \phi(x)}$
with arbitrary $\alpha$ is primary with scaling dimension $\alpha^2/2$.
The general correlation function is given by
\be
\langle e^{i\alpha\phi(z_1)} e^{-i\beta\phi(z_2)}\rangle=
e^{\alpha\beta G(z_1,z_2)-\alpha^2 G(z_1,z_1)/2 -\beta^2 G(z_2,z_2)/2}\,,
\ee
with $G(z_1,z_2)=\langle\phi(z_1) \phi(z_2) \rangle$.
Because of the restriction to the upper half-plane we have the Green function
\be
G(z_1,z_2)=G_\infty(z_1-z_2)- G_\infty(z_1-\bar{z}_2)\,
\ee
where $G_\infty(z)=G_\infty(0)-\log|z|$
is the Green function in the plane (we adsorbed the standard $\pi$ factors in
the definition of $\phi$) and we assumed a fixed boundary condition (free
ones are pathological because the one-point function is identically zero).
We then have
\be
\langle e^{i\alpha\phi(z_1)}e^{-i\beta\phi(z_2)}\rangle=
\frac1{|z_{1\bar1}|^{\alpha^2/2}|z_{2\bar2}|^{\beta^2/2}}
|z_{1\bar2}/z_{12}|^{\alpha\beta}\propto
\frac1{y_1^{\alpha^2/2}y_2^{\beta^2/2}} u^{2\alpha\beta}\,.
\ee
Using the previous formulas for the mapping to the strip and the
analytic continuation we have for $2t<|r_1-r_2|$ the independent evolution of
the two points, while for $2t>|r_1-r_2|$
\bea
\fl\langle e^{i\alpha\phi(z_1,t)}e^{-i\beta\phi(z_2,t)}\rangle=
\nonumber\\
\fl \qquad e^{-\pi\alpha^2t/4\tz}e^{-\pi\beta^2t/4\tz}
e^{\pi(2t-|r_1-r_2|)\alpha\beta/4\tz}=
e^{-\pi (\alpha^2+\beta^2- 2\alpha\beta)t/4\tz}
e^{-\pi\alpha\beta|r_1-r_2|/4\tz}\,,
\eea
that always decays exponentially to zero because
$(\alpha^2+\beta^2- 2\alpha\beta)>0$.

It is easy to show that this indeed is a general feature of the homogeneous
quench for different operators for any initial condition.
In fact the, two-point function of different operators $\Phi_1$ and $\Phi_2$
in the upper half-plane can be always written as (when
$\langle\Phi_1\rangle\neq0$ identically)
\be
\langle\Phi_1(1)\Phi_2(2)\rangle^{\mathtt U}=
\langle\Phi_1(1)\rangle^{\mathtt U}\langle\Phi_2(2)\rangle^{\mathtt U}
F(u^4)\,,
\ee
where $F(u^4)$ can be related to the hypergeometric function determining the four-point
functions in the bulk \cite{cftbook}.
However its complete expression is not required: we only need the asymptotic
behavior  of $F(u^4\simeq1)$ and $F(u^4\gg1)$.
From general scaling arguments, we have $F(1)=1$ (i.e. close to the boundary
the two-point function factorizes in one-points ones) and
$F(u^4\gg1)\sim u^{-4x_b}$. 
The exponent $x_b$ can be calculated for any pair of operators and any boundary
condition from the bulk four-point functions \cite{cftbook}.
Here, we only stress that if $\Phi_1=\Phi_2$ then $x_b= x_1=x_2$ to recover
the right behavior of the bulk two-point function, and in general
$x_b<(x_1+x_2)/2$ for unitarity.
Putting everything together we have
\be
\fl\langle\Phi_1(r_1,t)\Phi_2(r_2,0)\rangle=
\cases{
e^{-\pi x_1 t/2\tz}e^{-\pi x_2 t/2\tz} & $2t<|r_1-r_2|$,\cr
e^{-\pi x_1 t/2\tz}e^{-\pi x_2 t/2\tz} e^{\pi x_b (2t-|r_1-r_2|)/2\tz}
 & $2t>|r_1-r_2|$.
}
\ee
Note that when the operators $\Phi_1$ and $\Phi_2$ have the same scaling
dimensions the time dependence disappear and we get the usual thermal like
correlator. Oppositely, since $x_1+x_2-2x_b>0$, the correlation decays to zero
as anticipated.


\section*{References}

\begin{thebibliography}{99}

\bibitem{gv-01}
A. G\"orlitz, J. M. Vogels, A. E. Leanhardt, C. Raman, T. L. Gustavson,
J. R. Abo-Shaeer, A. P. Chikkatur, S. Gupta, S. Inouye, T. Rosenband
and W. Ketterle,
Realization of Bose-Einstein Condensates in Lower Dimensions,
Phys. Rev. Lett. {\bf 87}, 130402 (2001).

\bibitem{gb-01}
M. Greiner, I. Bloch, O. Mandel, T. W. H\"ansch, and T. Esslinger,
Exploring Phase Coherence in a 2D Lattice of Bose-Einstein Condensates,
Phys. Rev. Lett. {\bf 87}, 160405 (2001) [cond-mat/0105105].

\bibitem{mske-03}
H. Moritz, T. St\"oferle, M. K\"ohl, and T. Esslinger,
Exciting Collective Oscillations in a Trapped 1D Gas,
Phys. Rev. Lett. {\bf 91}, 250402 (2003) [cond-mat/0307607].

\bibitem{sm-04}
T. Stoferle, H. Moritz, C. Schori, M. Kohl, and T. Esslinger,
Transition from a strongly interacting 1D superfluid to a Mott insulator,
Phys. Rev. Lett. {\bf 92}, 130403 (2004) [cond-mat/0312440].

\bibitem{pw-04}
B. Paredes, A. Widera, V. Murg, O. Mandel, S. Falling,
I. Cirac, G. V. Shlyapnikov, T. W. Hansch, and I. Bloch,
Tonks-Girardeau gas of ultracold atoms in an optical lattice,
2004 Nature {\bf 429}, 277.

\bibitem{kwt-04}
T. Kinoshita, T. Wenger, and D. S. Weiss,
Observation of a One-Dimensional Tonks-Girardeau Gas,
2004 Science {\bf 305}, 1125.

\bibitem{kwt-05}
T. Kinoshita, T. Wenger and D. S. Weiss,
 Local Pair Correlations in One-Dimensional Bose Gases,
Phys. Rev. Lett. {\bf 95}, 190406 (2005).

\bibitem{aewkd-08}
A. H. van Amerongen, J. J. van Es, P. Wicke, K. V. Kheruntsyan, and N. J. van
Druten, Yang-Yang Thermodynamics on an Atom Chip,
Phys. Rev. Lett. {\bf 100}, 090402 (2008) [0709.1899].


\bibitem{bdw-07}
I. Bloch, J. Dalibard and W. Zwerger,
Many-Body Physics with Ultracold Gases,
Rev. Mod. Phys. to appear, [0704.3011].


\bibitem{uc}
M.~Greiner, O.~Mandel, T.~W.~H\"ansch, and I.~Bloch,
Collapse and Revival of the Matter Wave Field of a Bose-Einstein Condensate,
2002 Nature (London) {\bf 419}, 51 [cond-mat/0207196]

\bibitem{uc2}
C.~Orzel, A.~K.~Tuchman, M.~L.~Fenselau, M.~Yasuda, and
M.~A.~Kasevich, Squeezed States in a Bose-Einstein Condensate,
2001 Science {\bf 291} 2386.

\bibitem{uc3}
A. Widera, F. Gerbier, S. F\"olling, T. Gericke, O. Mandel, and I.
Bloch, Coherent Collisional Spin Dynamics in Optical Lattices, 2005
Phys. Rev. Lett. {\bf 95}, 190405.

\bibitem{spinor}
L. E. Sadler, J. M. Higbie, S. R. Leslie, M. Vengalattore,
and D. M. Stamper-Kurn,
Spontaneous symmetry breaking in a quenched ferromagnetic spinor Bose-Einstein
condensate,
2006 Nature {\bf 443}, 312.

\bibitem{kww-06}
T. Kinoshita, T. Wenger, and D. S. Weiss,
A quantum Newton's cradle,
2006 Nature {\bf 440}, 900.

\bibitem{wt-07}
A. Widera, S. Trotzky, P. Cheinet, S. Folling, F. Gerbier, I. Bloch,
V. Gritsev, M. D. Lukin, and E. Demler,
Quantum spin dynamics of squeezed Luttinger liquids in two-component atomic
gases, 0709.2094.

\bibitem{tc-07}
S. Trotzky, P. Cheinet, S. Fölling, M. Feld, U. Schnorrberger, A.
M. Rey, A. Polkovnikov, E. A. Demler, M. D. Lukin, and I. Bloch,
Time-resolved Observation and Control of Superexchange Interactions
with Ultracold Atoms in Optical Lattices,
Science {\bf 319}, 285 (2008) [0712.1853].

\bibitem{ao-03}
L. Amico and A. Osterloh,
Out of equilibrium correlation functions of quantum anisotropic XY models:
one-particle excitations,
2004 J. Phys. A {\bf 37} 291 [cond-mat/0306285].

\bibitem{aop-03}
L. Amico, A. Osterloh, F. Plastina, R. Fazio, and G. M. Palma,
Dynamics of Entanglement in One-Dimensional Spin Systems,
2004 Phys. Rev. A. {\bf 69}, 022304 [quant-ph/0307048].

\bibitem{sps-04}
K. Sengupta, S. Powell, and S. Sachdev,
Quench dynamics across quantum critical points,
2004 Phys. Rev. A {\bf 69} 053616 [cond-mat/0311355];\\
F. Igloi and H. Rieger,
Long-Range Correlations in the Nonequilibrium Quantum Relaxation
of a Spin Chain, 2000 Phys. Rev. Lett. {\bf 85} 3233 [cond-mat/0003193].

\bibitem{cc-05} P. Calabrese and J. Cardy,
Evolution of Entanglement entropy in one dimensional systems,
J. Stat. Mech. P04010 (2005) [cond-mat/0503393].

\bibitem{dmcf-06}
G. De Chiara, S. Montangero, P. Calabrese, and R. Fazio, Entanglement
Entropy dynamics in Heisenberg chains, 2006 J. Stat. Mech. P03001
[cond-mat/0512586].

\bibitem{cl-05}
R. W. Cherng and L. S. Levitov,
Entropy and Correlation Functions of a Driven Quantum Spin Chain,
2006 Phys. Rev. A {\bf 73}, 043614 [cond-mat/0512689].

\bibitem{mg-05}
A. Minguzzi and D.M. Gangardt,
Exact coherent states of a harmonically confined Tonks-Girardeau gas,
2005 Phys. Rev. Lett. {\bf 94}, 240404 [cond-mat/0504024].

\bibitem{cc-06} P. Calabrese and  J. Cardy,
Time-dependence of correlation functions following a quantum quench,
2006 Phys. Rev. Lett. {\bf 96}, 136801 [cond-mat/0601225].

\bibitem{gg} M. Rigol, V. Dunjko, V. Yurovsky, and M. Olshanii,
Relaxation in a Completely Integrable Many-Body Quantum System: An Ab Initio
Study of the Dynamics of the Highly Excited States of Lattice Hard-Core Bosons,
2007 Phys. Rev. Lett. {\bf 98}, 50405 [cond-mat/0604476].

\bibitem{c-06}
M. A. Cazalilla. Effect of suddenly turning on the interactions in
the Luttinger model, 2006 Phys. Rev. Lett. {\bf 97}, 156403
[cond-mat/0606236].

\bibitem{rmo-06}
M. Rigol, A. Muramatsu, and M. Olshanii,
Hard-core bosons on optical superlattices: Dynamics and relaxation in the
superfluid and insulating regimes,
2006 Phys. Rev. A {\bf 74}, 053616 [cond-mat/0612415].

\bibitem{cdeo-07}
M. Cramer, C.M. Dawson, J. Eisert, and T.J. Osborne,
Quenching, relaxation, and a central limit theorem for quantum lattice
systems, Phys. Rev. Lett. {\bf 100}, 030602 (2008)
[cond-mat/0703314].

\bibitem{gdlp-07}
V. Gritsev, E. Demler, M. Lukin, and A. Polkovnikov, Spectroscopy of
Collective Excitations in Interacting Low-Dimensional Many-Body
Systems Using Quench Dynamics, 2007 Phys. Rev. Lett. {\bf 99} 200404
[cond-mat/0702343].

\bibitem{cc-07} P. Calabrese and  J. Cardy,
 Quantum quenches in extended systems,
J. Stat. Mech. P06008 (2007) [0704.1880].

\bibitem{ep-07}
V. Eisler and I. Peschel, Evolution of entanglement after a local
quench, J. Stat. Mech. P06005 (2007) [cond-mat/0703379].

\bibitem{klr-07}
I. Klich, C. Lannert, and G. Refael,
Supercurrent survival under Rosen-Zener quench of hard core bosons,
Phys. Rev. Lett. {\bf 99}, 205303 (2007)  [0706.2869].

\bibitem{cc-07b}
P. Calabrese and  J. Cardy, Entanglement and correlation functions
following a local quench: a conformal field theory approach,
J. Stat. Mech. P10004 (2007) [0708.3750].

\bibitem{rdo-07}
M. Rigol, V. Dunjko, and M. Olshanii,
Thermalization and its mechanism for generic isolated quantum systems,
0708.1324.

\bibitem{bpz-07}
H. Buljan, R. Pezer, and T. Gasenzer,
Fermi-Bose transformation for the time-dependent Lieb-Liniger gas,
Phys. Rev. Lett. {\bf 100}, 080406 (2008) [0709.1444].

\bibitem{gp-07}
D. M. Gangardt and M. Pustilnik,
Correlations in an expanding gas of hard-core bosons, 0709.2374.



\bibitem{ekpp-07}
V. Eisler, D. Karevski, T. Platini, I. Peschel, Entanglement
evolution after connecting finite to infinite quantum chains,
J. Stat. Mech. (2008) P01023 [0711.0289].

\bibitem{bs-07}
T. Barthel and U. Schollwoeck,
Dephasing and the steady state in quantum many-particle systems,
Phys. Rev. Lett. {\bf 100}, 100601 (2008) [0711.4896].

\bibitem{brd-08}
P. Barmettler, A. M. Rey, E. Demler, M. D. Lukin, I. Bloch, and V. Gritsev,
Quantum Many-Body Dynamics of Coupled Double-Well Superlattices,
0803.1643.

\bibitem{qzm}
W. H. Zurek, U. Dorner, and P. Zoller,
Dynamics of a Quantum Phase Transition,
2005  Phys. Rev. Lett. {\bf 95} 105701 [cond-mat/0503511];\\
A. Polkovnikov,
Universal adiabatic dynamics across a quantum critical point,
2005 Phys. Rev. B {\bf 72}, 161201(R) [cond-mat/0312144];\\
J. Dziarmaga,
Dynamics of a quantum phase transition in the random Ising model,
2006 Phys. Rev. B {\bf 74}, 64416 [cond-mat/0603814];\\
L. Cincio, J. Dziarmaga, M. M. Rams, and W. H. Zurek,
Entropy of entanglement and correlations induced by a quench: Dynamics of a
quantum phase transition in the quantum Ising model,
Phys. Rev. A {\bf 75}, 052321 (2007) [cond-mat/0701768];\\
T. Caneva, R. Fazio, G. E. Santoro,
Adiabatic quantum dynamics of a random Ising chain across its quantum critical
point, Phys. Rev. B {\bf 76}, 144427 (2007) [0706.1832];\\
K. Sengupta, D. Sen, and S. Mondal,
Exact Results for Quench Dynamics and Defect Production in
a Two-Dimensional Model,
Phys. Rev. Lett. {\bf 100}, 077204 (2008) [0710.1712];\\
U. Divakaran and A. Dutta,
The effect of the three-spin interaction and the next nearest neighbor
interaction on the quenching dynamics of a transverse Ising model,
J. Stat. Mech.  (2007) P11001 [0801.0621];\\
F. Pellegrini, S. Montangero, G. E. Santoro, R. Fazio,
Adiabatic quenches through an extended quantum critical region,
0801.4475;\\
S. Mondal, D. Sen, and K. Sengupta,
Quench dynamics and defect production in the Kitaev and extended Kitaev
models,  0802.3986.

\bibitem{ep-08}
V. Eisler and I. Peschel, Entanglement in a periodic quench, 0803.2655.

\bibitem{tdmrg}
A. J. Daley, C. Kollath, U. Schollwoeck, and G. Vidal,
Time-dependent density-matrix renormalization-group using adaptive
effective Hilbert spaces,
J. Stat. Mech. P04005 (2004) [cond-mat/0403313];\\
S. R. White and A. E. Feiguin, Real time evolution using the density
matrix renormalization group, Phys. Rev. Lett. {\bf 93}, 076401
(2004) [cond-mat/0403310].

\bibitem{ksdz-05}
C. Kollath, U. Schollwoeck, J. von Delft, and W. Zwerger,
One-dimensional density waves of ultracold bosons in an optical lattice,
2005 Phys. Rev. A {\bf 71}, 053606 [cond-mat/0411403].

\bibitem{ksz-05}
C. Kollath, U. Schollwoeck, and W. Zwerger,
Spin-charge separation in cold Fermi-gases: a real time analysis,
2005 Phys. Rev. Lett. {\bf 95}, 176401 [cond-mat/0504299].

\bibitem{kla-06}
C. Kollath, A. Laeuchli, and E. Altman, Quench dynamics and non
equilibrium phase diagram of the Bose-Hubbard model, 2007 Phys. Rev.
Lett. 98, 180601 [cond-mat/0607235].

\bibitem{rmrnm-06}
K. Rodriguez, S.R. Manmana, M. Rigol, R.M. Noack, and A. Muramatsu,
Coherent matter waves emerging from Mott-insulators, 2006 New J.
Phys. {\bf 8}, 169 [cond-mat/0606155].

\bibitem{mwnm-06}
S.R. Manmana, S. Wessel, R.M. Noack, and A. Muramatsu, Strongly
correlated fermions after a quantum quench, 2007 Phys. Rev. Lett.
{\bf 98}, 210405 [cond-mat/0612030].

\bibitem{kkmgs-07}
A. Kleine, C. Kollath, I. P. McCulloch, T. Giamarchi, U.
Schollwoeck, Excitations in two-component Bose-gases,
0712.1448.

\bibitem{hrmfd-08}
F. Heidrich-Meisner, M. Rigol, A. Muramatsu, A.E. Feiguin, and E. Dagotto,
Ground-state reference systems for expanding correlated fermions in one
dimension, Phys. Rev. Lett., to appear [0801.4454].

\bibitem{lk-08}
A. Laeuchli and C. Kollath, 
Spreading of correlations and entanglement after a quench in the 
Bose-Hubbard model, 0803.2947.

\bibitem{arf-08}
K. A. Al-Hassanieh, F. A. Reboredo, A. E. Feiguin, I. Gonzalez, and 
E. Dagotto, Excitons in the One-Dimensional Hubbard Model: a Real-Time Study,
0804.0617.




\bibitem{dd}
H. W. Diehl, The theory of boundary critical phenomena, in
{\it Phase Transitions and Critical Phenomena}
vol 10 ed C. Domb and J. L. Lebowitz (1986, London: Academic)\\
H. W. Diehl, The theory of boundary critical phenomena,
Int. J. Mod. phys. B {\bf 11}, 3503 (1997) [cond-mat/9610143].

\bibitem{gc-08}
P. Calabrese and A. Gambassi,
Quantum quench close to a critical point: Landau-Ginzburg
order-parameter evolution, to appear.

\bibitem{kink}
T. Antal, Z. Racz, A. Rakos, and G. M. Schutz,
Transport in the XX chain at zero temperature: Emergence of flat magnetization
profiles, Phys. Rev. E {\bf 59}, 4912 (1999) [cond-mat/9812237];\\
D. Karevski, Scaling behaviour of the relaxation in quantum chains,
Eur. Phys. J. B {\bf 27}, 147 (2001) [cond-mat/0203078];\\
Y. Ogata, Diffusion of the Magnetization Profile in the XX-model,
Phys. Rev. E {\bf 66}, 066123 (2002) [cond-mat/0210011];\\
W. H. Aschbacher and  C.-A. Pillet,
Non-equilibrium steady states of the XY chain,
J. Stat. Phys. {\bf 112}, 1153 (2003);\\
T. Platini and D. Karevski,
Scaling and front dynamics in Ising quantum chains,
Eur. Phys. J. B {\bf 48}, 225 (2005) [cond-mat/0509594];\\
W. H. Aschbacher and J.-M. Barbaroux,
Out of equilibrium correlations in the XY chain,
Lett. Math. Phys. {\bf 77}, 11 (2006);\\
T. Platini and D. Karevski,
Relaxation in the XX quantum chain,
J. Phys. A {\bf 40}, 1711  (2007) [cond-mat/0611673];\\
W. H. Aschbacher,
Non-zero entropy density in the XY chain out of equilibrium,
Lett. Math. Phys. {\bf 79}, 1 (2007).



\bibitem{bx-90}
T. W. Burkhardt and T. Xue,
Conformal invariance and critical systems with mixed boundary conditions,
Nucl. Phys. B {\bf 354}, 653 (1990);\\
T. W. Burkhardt and T. Xue,
Density profiles in confined critical systems and conformal invariance,
Phys. Rev. Lett. {\bf 66}, 895 (1991).

\bibitem{c-84}
J. L. Cardy, Conformal Invariance and Surface Critical Behavior,
1984 Nucl. Phys. B {\bf 240} 514.

\bibitem{c-05} J. L. Cardy, Boundary Conformal Field Theory,
in {\sl Encyclopedia of Mathematical Physics}, ed J.-P. Francoise, G. Naber,
and S. Tsun Tsou, (Elsevier, Amsterdam, 2006).

\bibitem{6p}
T. W. Burkhardt and I. Guim,
Bulk, surface, and interface properties of the Ising model and conformal
invariance,
Phys. Rev. B {\bf 36}, 2080 (1987):\\
P. Di Francesco, H. Saleur, and J. B. Zuber,
 Critical Ising Correlation Functions In The Plane And On The Torus,
Nucl. Phys. B {\bf 290}, 527 (1987).





\bibitem{c-86}
J. L. Cardy, Effect of Boundary Conditions on the Operator Content of
Two-Dimensional Conformally Invariant Theories.
Nucl. Phys. B {\bf 275}, 200 (1986).

\bibitem{cl-91}
J. Cardy  and D. Lewellen, Bulk and Boundary Operators in Conformal
Field Theory, 1991 Phys. Lett. B {\bf 259} 274.


\bibitem{ent-rev}
L. Amico, R. Fazio, A. Osterloh, and V. Vedral,
Entanglement in Many-Body Systems,
Rev. Mod. Phys. to appear [quant-ph/0703044].

\bibitem{cc-04}
P.~Calabrese and J.~Cardy,
Entanglement entropy and quantum field theory,
J. Stat. Mech. P06002 (2004) [hep-th/0405152]

\bibitem{ee} C. Holzhey, F. Larsen, and F. Wilczek,
Geometric and Renormalized Entropy in Conformal Field Theory,
Nucl. Phys. B {\bf 424}, 443 (1994) [hep-th/9403108];\\
G. Vidal, J. I. Latorre, E. Rico, and A. Kitaev,
Entanglement in quantum critical phenomena,
Phys. Rev. Lett. {\bf 90}, 227902 (2003) [quant-ph/0211074]\\
J. I. Latorre, E. Rico, and G. Vidal,
Ground state entanglement in quantum spin chains,
Quant. Inf. and Comp. {\bf 4}, 048 (2004) [quant-ph/0304098].



\bibitem{eff-dmrg}
U. Schollwoeck, The density-matrix renormalization group,
Rev. Mod. Phys. {\bf 77}, 259 (2005) [cond-mat/0409292];\\
G. Vidal, Efficient simulation of one-dimensional quantum many-body systems,
Phys. Rev. Lett. {\bf 93}, 040502 (2004) [quant-ph/0310089];\\
T. J. Osborne,
The Dynamics of 1D Quantum Spin Systems Can Be Approximated Efficiently,
Phys. Rev. Lett. {\bf 97}, 157202 (2006) [quant-ph/0508031];\\
J. Eisert, Computational Difficulty of Global Variations in the Density Matrix
Renormalization Group,
Phys. Rev. Lett. {\bf 97}, 260501 (2006) [quant-ph/0609051];\\
N. Schuch, M. M. Wolf, F. Verstraete, and J. I. Cirac,
Entropy scaling and simulability by Matrix Product States,
Phys. Rev. Lett. {\bf 100}, 030504 (2008) [0705.0292];\\
A. Perales and G. Vidal,
Entanglement growth and simulation efficiency in one-dimensional quantum
lattice systems, 0711.3676;\\
N. Schuch, M. M. Wolf, K. G. H. Vollbrecht, and J. I. Cirac,
On entropy growth and the hardness of simulating time evolution,
New J. Phys. {\bf 10}, 033032 (2008) [0801.2078];\\
M. B. Hastings,
Observations Outside the Light-Cone: Algorithms for Non-Equilibrium and
Thermal States, 0801.2161.


\bibitem{sle-rev}
M. Bauer and D. Bernard,
2D growth processes: SLE and Loewner chains
Phys. Rep. {\bf 432} (2006) 115 [math-ph/0602049];\\
J. Cardy, SLE for theoretical physicists
Ann. Phys. {\bf 318} (2005) 81 [cond-mat/0503313];\\
W. Kager and B. Nienhuis,
A guide to stochastic Loewner evolution and its applications,
J. Stat. Phys. {\bf 115}, 1149 (2004) [math-ph/0312251];\\
I.A. Gruzberg and L.P. Kadanoff,
The Loewner equation: maps and shapes,
J. Stat. Phys. {\bf 114}, 1183 (2004) [cond-mat/0309292].

\bibitem{s-00}
O. Schramm,
Scaling limits of loop-erased random walks and uniform spanning trees,
Israel J. Math. {\bf 118}, 221, (2000) [math.PR/9904022].


\bibitem{gc-05}
A. Gamsa and J. Cardy,
The scaling limit of two cluster boundaries in critical lattice models,
J. Stat. Mech. (2005) P12009 [math-ph/0509004].

\bibitem{drc-05}
B. Doyon, V. Riva, and J. Cardy,
Identification of the stress-energy tensor through conformal restriction in
SLE and related processes,
Commun. Math. Phys. {\bf 268} (2006) 687.


\bibitem{bbk-05}
M. Bauer, D. Bernard, and K. Kytola,
Multiple Schramm-Loewner Evolutions and Statistical Mechanics Martingales,
J. Stat. Phys. {\bf 120}, 1125 (2005) [math-ph/0503024].

\bibitem{m-87}
M. P. Mattis, Correlation functions in 2-dimensional critical theories,
Nucl. Phys. B {\bf 285}, 671 (1987);\\
P. Arnold and M. P. Mattis, Operator products in 2-dimensional critical
theories, Nucl. Phys. B {\bf 295}, 363 (1988).

\bibitem{cftbook}
P. Di Francesco, P. Mathieu, and D. Senechal,
Conformal Field Theory (Springer-Verlag, New York, 1997).





\end{thebibliography}
\end{document}